\newif\ifHEVEA \newif\ifDRAFT \newif\ifFINAL \newif\ifREVWS \newif\ifEXPLE \newif\ifACCEPTED

\newif\ifTWOCOLS

\TWOCOLStrue

\HEVEAfalse
\DRAFTfalse
\REVWSfalse
\EXPLEfalse
\FINALtrue
\ACCEPTEDtrue

\TWOCOLStrue

\documentclass[sigconf]{acmart}

\newcommand{\abstractcount}[1]{}
\newcommand{\bodycount}{}

\ifHEVEA
\else

\fi

\usepackage[normalem]{ulem} 

\usepackage{xcolor}

\ifREVWS
  \newcommand{\reviews}[1]{\textcolor{orange}{#1}} \else
  \newcommand{\reviews}[1]{} \fi

\ifEXPLE
  \newcommand{\example}[1]{\textcolor{yellow}{#1}} \else
  \newcommand{\example}[1]{} \fi

\newcommand{\white}[1]{\textcolor{white}{#1}} \newcommand{\black}[1]{\textcolor{black}{#1}} 

\ifDRAFT
\newcommand{\barre}[1]{\textcolor{lightgray}{\sout{#1}}} \newcommand{\barrePT}[1]{\textcolor{lightgray}{\sout{#1}}} 

  \newcommand{\remark}[1]{}

         \newcommand{\purple}[1]{\textcolor{purple}{#1}} 

\else
  \newcommand{\barre}[1]{} \newcommand{\barrePT}[1]{} 

  \ifFINAL
    \renewcommand{\reviews}[1]{} \newcommand{\remark}[1]{}    \else
    \newcommand{\remark}[1]{\textcolor{red}{\textsc{#1}}}    \fi

        \newcommand{\purple}[1]{#1} \renewcommand{\textcolor}[2]{#2} \fi

\ifDRAFT
  \usepackage{soul}
  \definecolor{bleuclair}{rgb}{0.7, 0.7, 1.0}
  \definecolor{rosepale}{rgb}{1.0, 0.7, 1.0}

\else

\fi

\def\ie{{i.e.,} }
\def\eg{{e.g.,} }

\newcommand{\ita}[1]{\textit{#1}}

\newcommand{\itb}[1]{\textbf{\textit{#1}}}

\setcitestyle{notesep={: }}

\usepackage{eurosym}

\DeclareTextSymbol{\deg}{T1}{6}
\DeclareTextSymbol{\deg}{OT1}{23}

\newcommand\wordcount{
    \immediate\write18{texcount -sub=section \jobname.tex  | grep "Section" | sed -e 's/+.*//' | sed -n \thesection p > 'count.txt'}
(148
 words)}

\ifDRAFT
  \usepackage[yyyymmdd,24hr]{datetime} 
  \newcommand{\docversion}{<Document version: \today{} \currenttime{}>}
\else
  \newcommand\docversion[1]{}
\fi

\ifFINAL
  
\else
\fi

\usepackage{multirow}

\usepackage{tikz}
\usetikzlibrary{calc}

\makeatletter\let\expandableinput\@@input\makeatother
\usepackage{rotating} 

\definecolor{coulD}{RGB}{52,167,255} 
\definecolor{coulT}{RGB}{51,255,146} 
\definecolor{coulO}{RGB}{255,190,51} 
\definecolor{coulC}{RGB}{255,51,178}

\def\xd{\cellcolor{coulD}$\times$}
\def\xt{\cellcolor{coulT}$\times$}
\def\xo{\cellcolor{coulO}$\times$}
\def\xc{\cellcolor{coulC}$\times$}

\def\xs{\cellcolor{gray!95}$\times$} \def\xy{\cellcolor{gray!70}$\times$} \def\xl{\cellcolor{gray!50}$\times$}

\def\o{\cellcolor{lightgray!50}-}

\DeclareUnicodeCharacter{202F}{FIX ME!!!!}

\usepackage{colortbl}

\usepackage{subcaption}

\usepackage{enumitem}

\def\ie{{i.e.,} }
\def\eg{{e.g.,} }

\def\raL{\raggedleft}

\DeclareTextSymbol{\deg}{T1}{6}
\DeclareTextSymbol{\deg}{OT1}{23}

\ifHEVEA
  \providecommand{\tabularnewline}{\\}
  \newcommand{\url}{}

\else

\fi

\usepackage{amsmath}

\def\myshorttitle{A Framing and Analysis of Applicative Tangible Interfaces} \def\mytitle{A Framing and Analysis of Applicative Tangible Interfaces}

\def\mykeywords{tangible user interfaces,
tangible components,
applications,
interaction model,
taxonomy,
roles,
bodiness,
classification,
analysis
}


\AtBeginDocument{}

\setcopyright{cc}
\copyrightyear{2025}
\acmYear{2025}

\acmConference[]{arXiv.org, Cornell University}{2025}{Ithaca, NY, USA}
\acmPrice{}
\acmISBN{}

\begin{document}

\title[\myshorttitle]{\mytitle}

\author{Guillaume Rivière}
\email{g.riviere@estia.fr}
\orcid{0000-0001-8390-9751}
\affiliation{\institution{Univ. Bordeaux, ESTIA Institute of Technology}
  \city{Bidart}
  \postcode{F-64210}
  \country{France}
}

\renewcommand{\shortauthors}{Riviere}

\begin{abstract}

The investigation of tangible user interfaces commenced approximately thirty years ago.
Questions on its commercial potential become more pressing as the field becomes mature.
To take the field one step further---as the emergence of components contributed to the commercial development of graphical user interfaces---this article suggests that applicative tangible user interfaces could also be split into components.
These components are composed of the aggregation, combination, or coupling of physical items and fulfil four roles that are described through a new interaction model.
This article successfully distributed among these four components' roles all of the 159 physical items from a representative collection of 35 applications.
Further examination of these applicative tangible interfaces coincides with four research phases in the field and identifies three main paths for future research to fully realize the potential of tangible user interfaces.

\end{abstract}


\maketitle

\docversion{}

\abstractcount{150}

\bodycount{}

\newcommand\WT[1]{{\footnotesize \white{#1}}}
\newcommand\BT[1]{{\footnotesize \black{#1}}}
\newcommand\WB[1]{{\bf\footnotesize \white{#1}}}
\newcommand\BB[1]{{\bf\footnotesize \black{#1}}}
\newcommand\Wt[1]{{\scriptsize \white{#1}}}
\newcommand\Bt[1]{{\scriptsize \black{#1}}}

\newcommand\sbullet[1][.5]{\mathbin{\vcenter{\hbox{\scalebox{#1}{$\bullet$}}}}} 

\def\numslotmachine{\#1}
\def\numcaad{\#2}
\def\numselfbuilder{\#3}
\def\nummarble{\#4}
\def\numpassiveprops{\#5}
\def\numgraspdraw{\#6}
\def\nummetadesk{\#7}
\def\numbuildit{\#8}
\def\numpinwheels{\#9}
\def\nummediablocks{\#10}
\def\nummusicbottles{\#11}
\def\numurp{\#12}
\def\numsenseboard{\#13}
\def\numillclay{\#14}
\def\numaudiopad{\#15}
\def\numreactable{\#16}
\def\numipworkbench{\#17}
\def\numqueryshapes{\#18}
\def\numtuister{\#19}
\def\numiobrush{\#20}
\def\numpico{\#21}
\def\numchem{\#22}
\def\numnimio{\#23}
\def\numarcheotui{\#24}
\def\numgeotui{\#25}
\def\numslurp{\#26}
\def\numrelief{\#27}
\def\nummixitui{\#28}
\def\numteegi{\#29}
\def\numsoundform{\#30}
\def\numcairnform{\#31}
\def\numrespire{\#32}
\def\numaxes{\#33}
\def\numcoda{\#34}
\def\numsablier{\#35}

\def\numapps{35}
\def\numreps{159}
 
\newcounter{argument}
\setcounter{argument}{1}

\newcommand\ARG{\linebreak\linebreak\{\arabic{argument}\}\stepcounter{argument}}

\section{Introduction}

Since the conceptualization of tangible user interfaces (TUIs) in 1997 \cite{ishii1997tangible}, researchers have produced several hundred specimens over the past few decades. For instance, the Tangible Media Group at the MIT Media Lab alone generated an impressive 218 specimens\footnote{See the project page of the Tangible Media Group: \url{https://tangible.media.mit.edu/projects/} (last accessed on August 31, 2023).} between 1997 and 2023 (\ie{} an average of eight specimens yearly across twenty-six years). This landscape of tangible interfaces comprises prototypes showcasing new technologies (\eg{} The Actuated Workbench \cite{pangaro2002actuated}), and others exhibiting concrete applications of the concept (\eg{} Computational Optimization \cite{patten2007pico}). While the number of prototypes grew yearly, taxonomies, paradigms, and frameworks emerged to facilitate a better understanding of tangible user interfaces \citetext{\citealp{hoven2004tangible}, \citealp{mazalek2009framing}, \citealp[Ch.~5]{shaer2010past}, \citealp[Ch.~3]{ullmer2022weaving}}. In 2010 and, more recently, in 2022, an extensive review \cite{shaer2010past} and a thorough book \cite{ullmer2022weaving} provided comprehensive overviews of the field. With increasing maturity, there is growing scrutiny of the market viability and realism of tangible interfaces \cite{holmquist2019future,holmquist2023cheap}. When examining the history of graphical user interfaces (GUIs), the process of standardization and industrialization began with the imagination of the WIMP (Windows, Icons, Menus, Pointer) components\footnote{The origins of WIMP components date back to as early as 1968 or possibly even 1945 \cite{hiltzik2000dealers} (as cited in \cite{holmquist2023cheap}).} \cite{holmquist2023cheap,smith1982designing,smith1982overview} and their simplification \cite{holmquist2023cheap} in the early 1980s, which took advantage of the mouse invention over a decade earlier \cite{engelbart1968intellect}. Stimulated by this history, this article suggests studying the components found in the applicative tangible interfaces that have been proposed in the literature by raising the following questions. Could we split applicative tangible user interfaces into components? Are all these components required to develop applicative tangible user interfaces? Are there blind spots in tangible components that require further investigation before tangible interfaces could become ready for industrial development?

Previous work have framed the landscape of tangible user interfaces to better understand the degrees of tangibility and revealed trends in three same applicative domains to reach higher degrees of tangibility \cite{fishkin2004taxonomy}. This work frames applicative tangible user interfaces at the level of their tangible components, which are revealed from the aggregation of physical items into roles that are articulated through a new interaction model. Describing a collection of \numapps{} applicative tangible user interfaces (that deal with different application domains, or deal with same application domains differently) enables retrieving and confirming four development phases of the field, as well as scrutinizing underexplored opportunities of components for richer interaction and more complete applications, and providing insights for further exploring the components required for applicative tangible interfaces. Furthermore, these analyses are refined through four classes of bodiness that are defined for applicative tangible user interfaces according to the physical items involved in the implementation of their tangible data components.

First, this article provides an overview of the roles imagined in earlier models, notations, and toolkits, along with a summary of taxonomies and classifications regarding tangible interfaces. Subsequently, it establishes a novel taxonomy of physical items by relying on a new interaction model and constructs a classification of applicative tangible interfaces based on this taxonomy. It then showcases the taxonomy and the classification by detailing a collection of \numapps{} applications from the literature and examining interfaces' designs and the field's evolution. Finally, it concludes by addressing some future directions for the field and outlining some limitations of the work.

\section{Related Work}\label{sec-related-work}

Models of interaction and interfaces abstract the entities involved in interaction by assigning them roles. Then, interaction and interfaces can be described, understood, and compared across paradigms. Efforts to better understand tangible interfaces also call for taxonomies and classifications of these user interfaces. This section reviews (1) the roles described in some previous models, notations, and toolkits and (2) some taxonomies and classifications of tangible interfaces.

\subsection{Roles in Models, Notations, and Toolkits}\label{sec-related-models}

Instrumental interaction \cite{mbl2000instrumental} is an interaction model in which ``instruments'' are used to edit ``domain objects'' (\ie{} data role) attributes. These \textit{instruments} consist of a physical part (an input device) and a logical part (a graphical representation of the instrument) that translate user actions into commands that edit domain objects. Instruments provide the user with reactions from the instrument on its actions and feedback from edited object responses. In time-multiplexed conditions, activating the instrument entails associating the physical and logical parts \cite{mbl2000instrumental}. For example, the instrument may be activated by directly pointing at it or clicking a button in a tool palette \cite{mbl2000instrumental}. Instrumental interaction permits the description and comparison of WIMP interaction and interaction with graspable user interfaces \cite{fitzmaurice1995bricks} within a unified interaction model \cite{mbl2000instrumental}.

In the mixed interaction model \cite{coutrix2006mixed}, instruments are refined as ``mixed tools'' (\ie{} tools role), and domain objects become ``task objects'' (\ie{} data role), both of which are categorized as ``mixed objects.'' \textit{Mixed objects} possess physical and logical properties that are interconnected through ``linking modalities'' \cite{coutrix2006mixed}. \ita{Input linking modalities} sense physical proprieties using input devices and convert them into digital properties through \ita{input languages} \cite{coutrix2006mixed}. Conversely, \ita{output linking modalities} utilize \ita{output languages} and devices to translate digital properties into physical properties \cite{coutrix2006mixed}. Hence, where instrumental interaction deals only with domain objects shown graphically (on monitors or tabletops), mixed interaction extends to ``bodied task objects'' (\ie{} bodied data) that can provide input and output at the same time.

ASUR++ \cite{dubois2002asur++,dubois2003asur++} is an extension of the ASUR notation \cite{dubois2002asur} that characterizes the physical and digital entities of mobile mixed systems (\ie{} Adapter, System, User, and Real object). The notation is also suitable for tangible user interfaces \cite{dubois2003asur++}. The notation distinguishes between two digital roles for physical objects by defining terms for \ita{tools} (``R\textsubscript{tool}'') that aid in a task (\ie{} tools role) and \ita{targeted objects} (``R\textsubscript{object}'') that are altered by a task (\ie{} data role).

The ROSS API \cite{wu2012api}---Responsive Objects, Surfaces, and Spaces---outlines four nested levels for tangible interface elements, including the description of physical objects (``RObject''), as well as active spaces (``RSpace'') and surfaces (``RSurface'') capable of sensing interaction. Thus, a key advantage of this low-level framework is its ability to go beyond physical objects' description. However, even if input controls such as buttons, switches, and sliders are described (``RControl''), the API does not include descriptions of the roles of physical objects such as data and tools. Nonetheless, the toolkit's description of surfaces---which indicate where physical objects are placed and where graphics are displayed and touched, such as tabletop surfaces and mobile phone screens---suggests that limiting to the two roles of data and tools may be insufficient for fully describing interaction with tangible interfaces.

Data and tools also feature in tangible interfaces' paradigms \cite{holmquist1999token,shaer2004tac,ullmer2002phdthesis,ullmer2005token}, in addition to operations and constraints. In the Token-Based Access to Digital Information \cite{holmquist1999token}, ``tokens'' are bodied items of specific form to manipulate digital information (\ie{} data role) within applications, where as ``containers'' are bodied items of generic form to move any information between devices and applications, ``faucets'' are intangible items that convey digital information, and ``tools'' are bodied items bound to any computational functions without distinction between data editing (\eg{} physical handles to scale graphical shapes \cite{fitzmaurice1995bricks}) and application features (\eg{} applying filters to data rendering with the passive lens \cite{ishii1997tangible}). In the Token+Constraint approach \cite{ullmer2002phdthesis,ullmer2005token}, ``tokens'' are bodied items that represent digital information (\ie{} data role) and ``constraints'' are visually or mechanically delimited regions (\ie{} constraints role) that are bound to computational operations (\eg{} show, save). These operations are triggered when a token is sensed for presence, translation, or rotation within the constrained region. In the TAC Paradigm \cite{shaer2004tac}, ``tokens'' refer to the physical representations of digital information (\ie{} data role) or computational functions (\ie{} tools role), and ``constraints'' refer to the physical items that structure the user's manipulation of the tokens (\ie{} constraints role).

Thereby, the different roles that are found in tangible user interfaces can be summarized as data, tools, operations, and constraints. However, interaction models, notations, toolkits, and paradigms describe physical items only in the range of data, tools, and constraints (see \autoref{tab::previous-roles}). Operations with the application (\eg{} show, browse, color mode) or the operating system (\eg{} load, save, authentication) are melted with tools that edit data (\eg{} in \cite{holmquist1999token}) or bound to some trigger conditions between physical items such as tokens and constraints \cite{shaer2004tac,ullmer2002phdthesis,ullmer2005token}. Nevertheless, the tangible components assigned to the operations' role have later been coined and defined as ``core tangibles'' and ``domain tangibles'' \cite{ullmer2008core}, which are tangible components that implement operations shared among multiple applications (\eg{} load, save, authentication) or some application domains (\eg{} setting color in drawing applications), regardless of the application's interaction style \cite{ullmer2008core}.

\begin{table*}
{\centering\small
  \begin{tabular}{
  lll
  @{\hspace{4pt}}m{0mm}
  llll
  }
\midrule

\multicolumn{3}{l}{\itb{Previous Work}} && \multicolumn{4}{l}{\itb{Components' Roles}}  \tabularnewline

\cmidrule{1-3}\cmidrule{5-8}

\ita{Name} & \ita{Year} & \ita{References}  && \ita{Data} & \ita{Tools} & \ita{Operations} & \ita{Constraints} \tabularnewline

\midrule

Instrumental Interaction & 1997 & \cite{mbl1997instrumental,mbl2000instrumental} && Domain objects              & Instruments           & - & -                \tabularnewline

Token-Based Access       & 1999 & \cite{holmquist1999token}                      && Containers, Tokens, Faucets & Tools                 & - & -                \tabularnewline

Token+Constraint         & 2002 & \cite{ullmer2002phdthesis,ullmer2005token}     && Tokens                      & -                     & - & Constraints      \tabularnewline

ASUR++                   & 2002 & \cite{dubois2002asur++,dubois2003asur++}       && R\textsubscript{object}     & R\textsubscript{tool} & - & -                \tabularnewline

TAC Paradigm             & 2004 & \cite{shaer2004tac}                            && Tokens                      & Tokens                & - & Constraints      \tabularnewline

Mixed Interaction        & 2006 & \cite{coutrix2006mixed}                        && Task objects                & Mixed tools           & - & -                \tabularnewline

\midrule
  \end{tabular}}
\caption{Interface components' roles among some models, notations, and paradigms in tangible and mixed reality scopes.}~\label{tab::previous-roles}
\end{table*}

To expound upon these models, notations, and paradigms, this article relies on the four roles of data, tools, operations, and constraints to introduce a role-based interaction model, and then a taxonomy and a classification based on these roles.

\subsection{Taxonomies and Classifications}\label{sec-related-taxonomies}

\def\tuigenresfootnotetext{Eight genres of tangible user interface applications were described in 2008 \cite{ishii2008beyond}: (1)~tangible telepresence, (2)~tangibles with kinetic memory, (3)~constructive assembly, (4)~tokens and constraints, (5)~interactive surfaces---tabletop TUI, (6)~continuous plastic TUI, (7)~augmented everyday objects, and (8)~ambient media.}

The literature has utilized taxonomies and classifications to chart the conceptual space of tangible user interfaces. The seminal work on graspable user interfaces presented a conceptual space with thirteen axes \cite{fitzmaurice1995bricks}. Later, Ishii categorized tangible user interfaces into eight genres \cite{ishii2008beyond}. Various frameworks---approximately two dozen in total---have organized the conceptual space along different axes \citetext{\citealp{mazalek2009framing}, \citealp[Ch.~5]{shaer2010past}, \citealp[Ch.~3]{ullmer2022weaving}}. Those frameworks organized tangible interfaces by abstracting, designing, or building their experiences, domains, physicality, interactions, or technologies \cite{mazalek2009framing}. For instance, Fishkin's taxonomy of tangibility \cite{fishkin2004taxonomy} has organized tangible interfaces as a conceptual space of twenty categories along two axes: \ita{embodiment} (\ie{} the physical relationship between the input focus and the output focus of a tangible interface) and \ita{metaphor} (\ie{} the analogy between user actions or artifact meaning compared with real-world actions or objects, respectively). The analysis of nine applicative tangible interfaces---that were selected because dealing with the same three task domains among over 60 tangible user interfaces---revealed a trend toward seeking a higher degree of tangibility \cite{fishkin2004taxonomy}. The next section develops a new taxonomy that categorizes the physical items that constitute tangible interface components.

\section{Proposing a Taxonomy and a Classification}\label{sec-taxonomy}

This section provides a taxonomy of the physical items that are utilized in tangible user interface's components, based on their bodiness, by relying on four tangible-component roles found in applicative tangible interfaces, and introduces a new interaction model based on the relationships between these roles, before it splits applicative tangible interfaces into four bodiness classes.

The concept of a ``tangible user interface'' is considered here broadly to include any user interface that strives for a ``\textit{seamless coupling between bits and atoms}'' using ``atoms'' as representations and controls of digital information and functions \cite{ishii1997tangible}. Such tangible and related physical user interfaces range from ``passive real-world props'' \cite{hinckley1994passiveprops} to ``graspable user interfaces'' \cite{fitzmaurice1995bricks}, ``manipulative user interfaces'' \cite{harrison1998manipulative}, and ``embodied user interfaces'' \cite{fishkin1999embodied}. Without a difference, those approaches were simply advancements within a developing field \cite{fishkin2004taxonomy}, which combined in 1997 under the cohesive vision of ``tangible bits'' \cite{ishii1997tangible}.   Furthermore, in this article, the term ``bodied'' replaces the term ``physical'' when referring to some graspable elements; then the term ``physical'' encompasses both graspable and non-graspable entities (such as graphics, video projection, and audio).

However, before introducing the two axes of the taxonomy, this section priorly requires a clarification of the three following concepts: entities, components, and items.

\subsection{Entities, Components, and Items}

Let us first define the set of the physical interactional entities that are involved in tangible and related physical user interfaces. ``Physical Interactional Entities'' are considered the highest level of abstraction of anything that is generated, animated, or sensed by computing resources and that is perceivable, editable, or manipulable by users. Such ``Interactional Entities'' are composable, couplable, and hybridable. ``Physical'' comprises matter (\eg{} objects, sides, and substances), sound (\eg{} audio, percussion, and noise), and any graphical item (\eg{} shapes, texts, and images). For example, the sides of a cube can be considered as entities, as well as the whole cube, depending on what is sensed, mapped, or represented. Some other entities are added to this initial set, which are not generated, nor animated, nor sensed by computing resources, but which support, hold, or guide entities from the initial set during interaction (\eg{} surfaces, slots, and trails).

The term ``component'' refers to the entities that fulfil roles in user interaction, such as windows, icons, menus, and pointers components in the WIMP paradigm for graphical user interfaces. In addition, components are made of ``items.'' As an illustration, in the WIMP paradigm, window components basically consist of items such as an outer border, header bar, title, close button, and content pane. Likewise, desktops' icon components consist of two items: a pictogram and a caption.

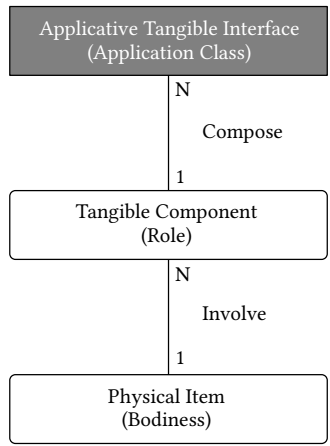
\begin{figure}\centering
  \def\svgwidth{\ifTWOCOLS 0.50\columnwidth \else 0.25\columnwidth \fi}
  {\small \begingroup \makeatletter \providecommand\color[2][]{\errmessage{(Inkscape) Color is used for the text in Inkscape, but the package 'color.sty' is not loaded}\renewcommand\color[2][]{}}\providecommand\transparent[1]{\errmessage{(Inkscape) Transparency is used (non-zero) for the text in Inkscape, but the package 'transparent.sty' is not loaded}\renewcommand\transparent[1]{}}\providecommand\rotatebox[2]{#2}\newcommand*\fsize{\dimexpr\f@size pt\relax}\newcommand*\lineheight[1]{\fontsize{\fsize}{#1\fsize}\selectfont}\ifx\svgwidth\undefined \setlength{\unitlength}{283.46456693bp}\ifx\svgscale\undefined \relax \else \setlength{\unitlength}{\unitlength * \real{\svgscale}}\fi \else \setlength{\unitlength}{\svgwidth}\fi \global\let\svgwidth\undefined \global\let\svgscale\undefined \makeatother \begin{picture}(1,1.3758609)\lineheight{1}\setlength\tabcolsep{0pt}\put(0,0){\includegraphics[width=\unitlength,page=1]{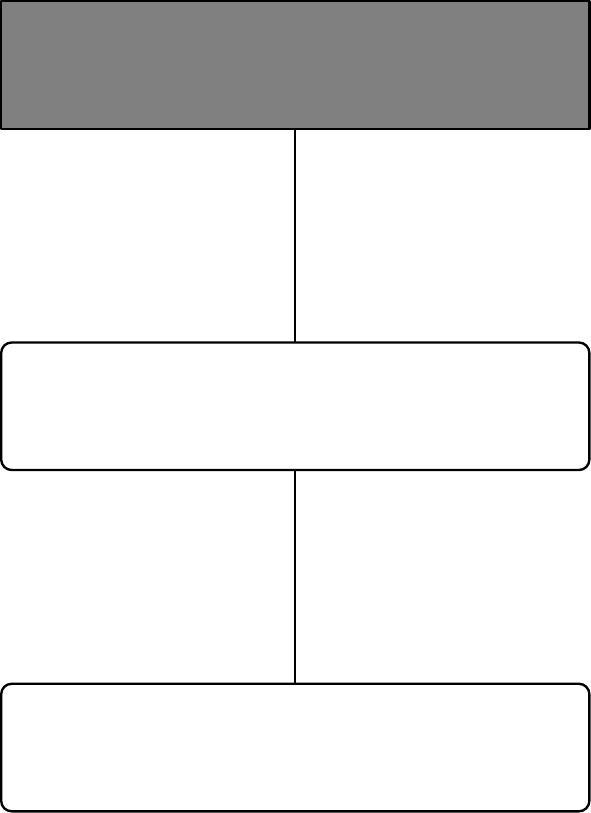}}\put(0.52166896,0.8192771){\makebox(0,0)[lt]{\lineheight{1.25}\smash{\begin{tabular}[t]{l}1\end{tabular}}}}\put(0.52166896,1.09444756){\makebox(0,0)[lt]{\lineheight{1.25}\smash{\begin{tabular}[t]{l}N\end{tabular}}}}\put(0.52166896,0.25306029){\makebox(0,0)[lt]{\lineheight{1.25}\smash{\begin{tabular}[t]{l}1\end{tabular}}}}\put(0.52166896,0.51235572){\makebox(0,0)[lt]{\lineheight{1.25}\smash{\begin{tabular}[t]{l}N\end{tabular}}}}\put(0.60675315,0.96132384){\makebox(0,0)[lt]{\lineheight{1.25}\smash{\begin{tabular}[t]{l}Compose\end{tabular}}}}\put(0.60675315,0.38902961){\makebox(0,0)[lt]{\lineheight{1.25}\smash{\begin{tabular}[t]{l}Involve\end{tabular}}}}\put(0.50290763,1.28563934){\color[rgb]{1,1,1}\makebox(0,0)[t]{\lineheight{0.94999999}\smash{\begin{tabular}[t]{c}Applicative Tangible Interface\end{tabular}}}}\put(0.50290765,1.21063935){\color[rgb]{1,1,1}\makebox(0,0)[t]{\lineheight{0.94999999}\smash{\begin{tabular}[t]{c}(Application Class)\end{tabular}}}}\put(0.49408425,0.70859085){\color[rgb]{0,0,0}\makebox(0,0)[t]{\lineheight{1.25}\smash{\begin{tabular}[t]{c}Tangible Component\end{tabular}}}}\put(0.49408425,0.63359085){\color[rgb]{0,0,0}\makebox(0,0)[t]{\lineheight{1.25}\smash{\begin{tabular}[t]{c}(Role)\end{tabular}}}}\put(0.49408424,0.13066043){\color[rgb]{0,0,0}\makebox(0,0)[t]{\lineheight{1.25}\smash{\begin{tabular}[t]{c}Physical Item\end{tabular}}}}\put(0.49408425,0.05566043){\color[rgb]{0,0,0}\makebox(0,0)[t]{\lineheight{1.25}\smash{\begin{tabular}[t]{c}(Bodiness)\end{tabular}}}}\end{picture}\endgroup  }
  \caption{The hierarchy of the elements that compose applicative tangible interfaces.}
  \Description{This diagram depicts three linked boxes, which entail applicative tangible interfaces, components, and items. The first link indicates that applicative tangible interfaces employ one or more components, and that components are employed by one application. The second link indicates that components involve one or more items, and that items are involved in one component.}
  \label{fig:components}
\end{figure}

In applicative tangible user interfaces, an interactive role in the user dialogue with computing resources can be insured by a single interactional entity itself, or by several interactional entities that are aggregated, combined, or coupled to fulfil this whole role. Thereby, a ``tangible component'' is such an interactional entity that fulfil an interactive role, or a subset of the interactional entities that fulfil an interactive role together. The ``physical items'' of a tangible user interface are any one-piece interactional entity whose bodiness can be defined and that are involved into interactive roles. Those physical items comprise elements that are sensed or controlled by computing resources, as well as some that are not (\ie{} even if they do not directly represent digital information or functions, \eg{} when insuring some mechanical or stuctural functions). As a consequence, a tangible component can aggregate, combine, or couple several physical items (of same or different bodinesses) to fulfil its role (\eg{} a bodied cursor and a bodiless value are coupled to constitute the color map amplitude tool of Teegi \cite{frey2014teegi}), as well as a tangible component can be composed of a single physical item. However, the number of physical items associated to the hierarchy of a component can also depend on the description's level of granularity (such as with the nested levels of the ROSS API \cite{wu2012api}). This hierarchy is summarized in \autoref{fig:components}.

The next subsection details the first axis of the taxonomy, which delineates four roles for tangible interface components.

\subsection{Four Components' Roles}

\begin{figure}\centering
  \def\svgwidth{\ifTWOCOLS \columnwidth \else 0.5\columnwidth \fi}
  {\small \begingroup \makeatletter \providecommand\color[2][]{\errmessage{(Inkscape) Color is used for the text in Inkscape, but the package 'color.sty' is not loaded}\renewcommand\color[2][]{}}\providecommand\transparent[1]{\errmessage{(Inkscape) Transparency is used (non-zero) for the text in Inkscape, but the package 'transparent.sty' is not loaded}\renewcommand\transparent[1]{}}\providecommand\rotatebox[2]{#2}\newcommand*\fsize{\dimexpr\f@size pt\relax}\newcommand*\lineheight[1]{\fontsize{\fsize}{#1\fsize}\selectfont}\ifx\svgwidth\undefined \setlength{\unitlength}{663.28667108bp}\ifx\svgscale\undefined \relax \else \setlength{\unitlength}{\unitlength * \real{\svgscale}}\fi \else \setlength{\unitlength}{\svgwidth}\fi \global\let\svgwidth\undefined \global\let\svgscale\undefined \makeatother \begin{picture}(1,1.20961187)\lineheight{1}\setlength\tabcolsep{0pt}\put(0,0){\includegraphics[width=\unitlength,page=1]{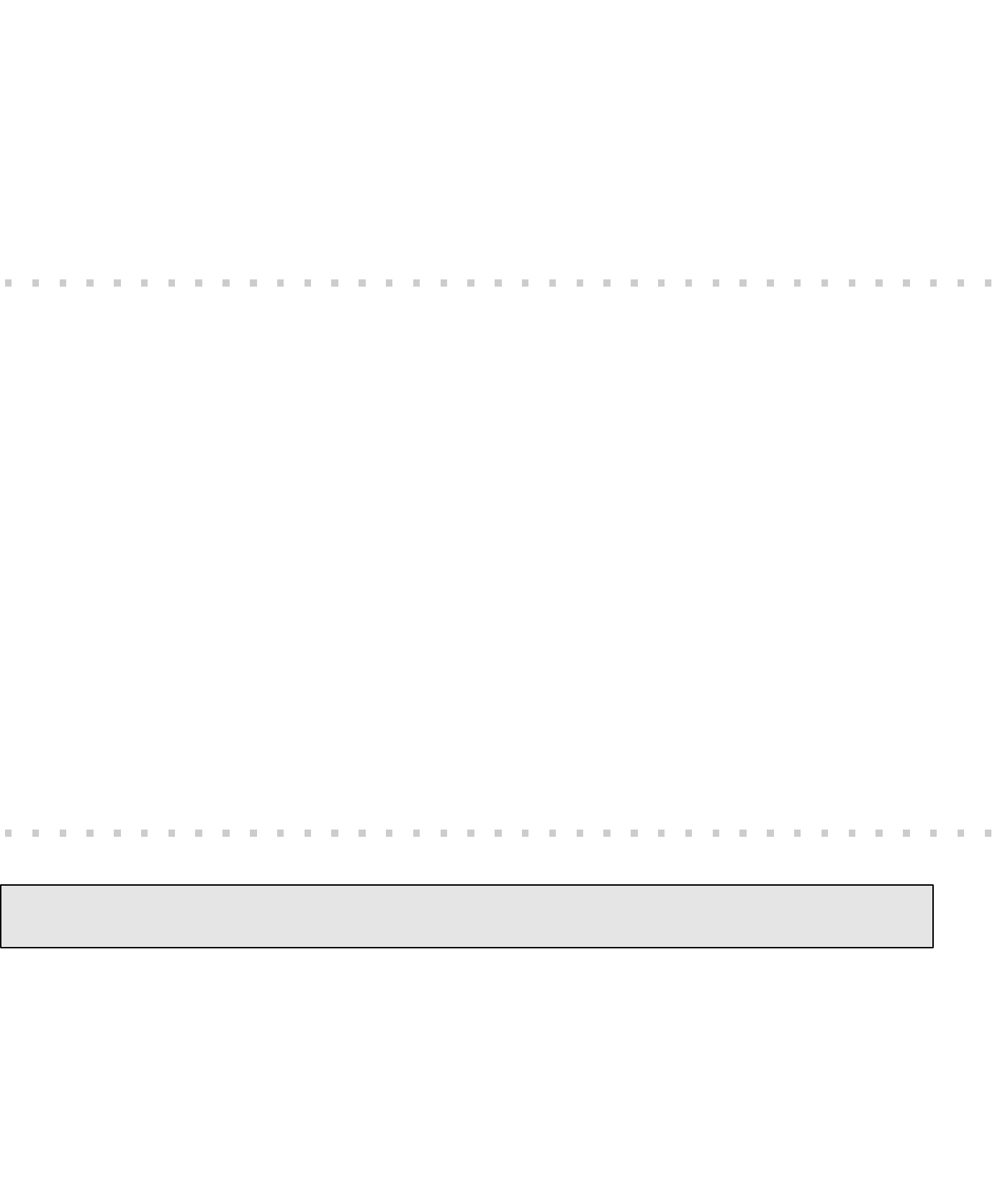}}\put(0.85593259,0.6405735){\rotatebox{90}{\makebox(0,0)[lt]{\lineheight{1.25}\smash{\begin{tabular}[t]{l}Feeds\end{tabular}}}}}\put(0.22665106,0.85056095){\makebox(0,0)[lt]{\lineheight{1.25}\smash{\begin{tabular}[t]{l}Load, Save, Browse...\end{tabular}}}}\put(0.45529071,0.57123445){\rotatebox{90}{\makebox(0,0)[lt]{\lineheight{1.25}\smash{\begin{tabular}[t]{l}Frame\end{tabular}}}}}\put(0.20221558,0.70382149){\rotatebox{-45}{\makebox(0,0)[lt]{\lineheight{1.25}\smash{\begin{tabular}[t]{l}Frame (+ load,\\save, play...)\end{tabular}}}}}\put(0.57983787,0.54598729){\rotatebox{45}{\makebox(0,0)[lt]{\lineheight{1.25}\smash{\begin{tabular}[t]{l}Frame (+ trigger)\end{tabular}}}}}\put(0,0){\includegraphics[width=\unitlength,page=2]{roles6_coul2.pdf}}\put(0.33849038,0.00524755){\makebox(0,0)[lt]{\lineheight{1.25}\smash{\begin{tabular}[t]{l}\BT{Role}\end{tabular}}}}\put(0.23686299,0.77642525){\makebox(0,0)[lt]{\lineheight{1.25}\smash{\begin{tabular}[t]{l}Edit\end{tabular}}}}\put(0.00090813,0.09161182){\makebox(0,0)[lt]{\lineheight{1.25}\smash{\begin{tabular}[t]{l}\BT{Legend:}\end{tabular}}}}\put(0,0){\includegraphics[width=\unitlength,page=3]{roles6_coul2.pdf}}\put(0.08996317,0.00578932){\makebox(0,0)[lt]{\lineheight{1.25}\smash{\begin{tabular}[t]{l}\BT{Program}\end{tabular}}}}\put(0.01673841,0.04638299){\makebox(0,0)[lt]{\lineheight{1.25}\smash{\begin{tabular}[t]{l}\textit{\BT{Digital Entities}}\end{tabular}}}}\put(0.26959854,0.04649515){\makebox(0,0)[lt]{\lineheight{1.25}\smash{\begin{tabular}[t]{l}\textit{\BT{Components}}\end{tabular}}}}\put(0.47653551,0.04562681){\makebox(0,0)[lt]{\lineheight{1.25}\smash{\begin{tabular}[t]{l}\textit{\BT{Relations}}\end{tabular}}}}\put(0.59432062,0.737904){\makebox(0,0)[lt]{\lineheight{1.25}\smash{\begin{tabular}[t]{l}Adjust\end{tabular}}}}\put(0,0){\includegraphics[width=\unitlength,page=4]{roles6_coul2.pdf}}\put(0.5868408,1.16871182){\rotatebox{-10}{\makebox(0,0)[lt]{\lineheight{1.25}\smash{\begin{tabular}[t]{l}Inform\end{tabular}}}}}\put(0.80298265,0.97419655){\rotatebox{-75}{\makebox(0,0)[lt]{\lineheight{0.75}\smash{\begin{tabular}[t]{l}Acts on\end{tabular}}}}}\put(0.591589,1.04701524){\rotatebox{-25}{\makebox(0,0)[lt]{\lineheight{0.94999999}\smash{\begin{tabular}[t]{l}Feed back\end{tabular}}}}}\put(0,0){\includegraphics[width=\unitlength,page=5]{roles6_coul2.pdf}}\put(0.77781144,0.04143156){\makebox(0,0)[lt]{\lineheight{1.25}\smash{\begin{tabular}[t]{l}\BT{Interactions}\end{tabular}}}}\put(0,0){\includegraphics[width=\unitlength,page=6]{roles6_coul2.pdf}}\put(0.77781144,0.00524755){\makebox(0,0)[lt]{\lineheight{1.25}\smash{\begin{tabular}[t]{l}\BT{Reactions}\end{tabular}}}}\put(0.94412863,0.6405735){\rotatebox{90}{\makebox(0,0)[lt]{\lineheight{1.25}\smash{\begin{tabular}[t]{l}Feeds\end{tabular}}}}}\put(0.73330879,0.33432846){\makebox(0,0)[lt]{\lineheight{1.25}\smash{\begin{tabular}[t]{l}Call\end{tabular}}}}\put(0.8237662,0.22351814){\makebox(0,0)[lt]{\lineheight{1.25}\smash{\begin{tabular}[t]{l}Call\end{tabular}}}}\put(0,0){\includegraphics[width=\unitlength,page=7]{roles6_coul2.pdf}}\put(0.47024842,0.17486386){\makebox(0,0)[t]{\lineheight{0.94999999}\smash{\begin{tabular}[t]{c}Operating System\end{tabular}}}}\put(0.4513771,0.9887982){\makebox(0,0)[lt]{\lineheight{0.94999999}\smash{\begin{tabular}[t]{l}Feed\\back\end{tabular}}}}\put(0.55608972,0.80038616){\makebox(0,0)[lt]{\lineheight{1.25}\smash{\begin{tabular}[t]{l}Acts on\end{tabular}}}}\put(0,0){\includegraphics[width=\unitlength,page=8]{roles6_coul2.pdf}}\put(0.47007001,0.72979687){\color[rgb]{0,0,0}\makebox(0,0)[t]{\lineheight{1.25}\smash{\begin{tabular}[t]{c}Tools\end{tabular}}}}\put(0,0){\includegraphics[width=\unitlength,page=9]{roles6_coul2.pdf}}\put(0.10542927,0.79246821){\color[rgb]{0,0,0}\makebox(0,0)[t]{\lineheight{1.25}\smash{\begin{tabular}[t]{c}Data\end{tabular}}}}\put(0,0){\includegraphics[width=\unitlength,page=10]{roles6_coul2.pdf}}\put(0.83440841,0.79443522){\color[rgb]{0,0,0}\makebox(0,0)[t]{\lineheight{1.25}\smash{\begin{tabular}[t]{c}Operations\end{tabular}}}}\put(0,0){\includegraphics[width=\unitlength,page=11]{roles6_coul2.pdf}}\put(0.28613076,1.15151142){\rotatebox{10}{\makebox(0,0)[lt]{\lineheight{1.25}\smash{\begin{tabular}[t]{l}Inform\end{tabular}}}}}\put(0.12832974,0.88875697){\rotatebox{75}{\makebox(0,0)[lt]{\lineheight{0.75}\smash{\begin{tabular}[t]{l}Acts on\end{tabular}}}}}\put(0.24771147,0.99370759){\rotatebox{25}{\makebox(0,0)[lt]{\lineheight{0.94999999}\smash{\begin{tabular}[t]{l}Feed back\end{tabular}}}}}\put(0.9997588,1.00297622){\color[rgb]{0.8,0.8,0.8}\rotatebox{90}{\makebox(0,0)[lt]{\lineheight{1.25}\smash{\begin{tabular}[t]{l}\textbf{\textit{HUMAN}}\end{tabular}}}}}\put(0.96960585,0.45373301){\color[rgb]{0.8,0.8,0.8}\rotatebox{90}{\makebox(0,0)[lt]{\lineheight{1.25}\smash{\begin{tabular}[t]{l}\textbf{\textit{TANGIBLE}}\\\textbf{\textit{INTERFACE}}\end{tabular}}}}}\put(0.97684252,0.15037927){\color[rgb]{0.8,0.8,0.8}\rotatebox{90}{\makebox(0,0)[lt]{\lineheight{0.94999999}\smash{\begin{tabular}[t]{l}\textbf{\textit{COMPUTING}}\\\textbf{\textit{RESOURCES}}\end{tabular}}}}}\put(0.47098414,0.28211532){\makebox(0,0)[t]{\lineheight{0.94999999}\smash{\begin{tabular}[t]{c}Application\\\end{tabular}}}}\put(0,0){\includegraphics[width=\unitlength,page=12]{roles6_coul2.pdf}}\put(0.4686449,0.47019931){\color[rgb]{0,0,0}\makebox(0,0)[t]{\lineheight{1.25}\smash{\begin{tabular}[t]{c}Constraints\end{tabular}}}}\put(0.50504228,0.55088154){\rotatebox{90}{\makebox(0,0)[lt]{\lineheight{1.25}\smash{\begin{tabular}[t]{l}(+ adjust)\end{tabular}}}}}\put(0,0){\includegraphics[width=\unitlength,page=13]{roles6_coul2.pdf}}\put(0.55845223,0.00672691){\makebox(0,0)[lt]{\lineheight{1.25}\smash{\begin{tabular}[t]{l}\BT{Messages}\end{tabular}}}}\put(0,0){\includegraphics[width=\unitlength,page=14]{roles6_coul2.pdf}}\put(0.77781705,0.07755035){\makebox(0,0)[lt]{\lineheight{1.25}\smash{\begin{tabular}[t]{l}\BT{Spatial concordance}\end{tabular}}}}\put(0.44727849,1.07532287){\makebox(0,0)[lt]{\lineheight{1.25}\smash{\begin{tabular}[t]{l}User\end{tabular}}}}\end{picture}\endgroup  }
\caption{The DTOC interaction model of tangible components' roles in applicative tangible interfaces.}
\Description{This diagram depicts the relations between four distinct roles of physical representations in tangible interfaces: data, tools, operations, and constraints. Operations adjust tools and customize data. Tools edit data. Constraints frame data, tools, and operations when spatial concordance, and are eventually assigned to supplementary commands that edit data, adjust tools, or trigger operations. Operations call the interactive application or the operating system. The interactive application and the operating system feed operations. The user acts on data, tools, and operations and receives feedback from them. The user gains information from both data and operations.}
  \label{fig:roles}
\end{figure}

The first axis concerns components' roles in which physical items are involved. This axis comprises four values drawn from prior framings of tangible interfaces, mixed reality, and instruments: data, tools, operations, and constraints. The relationships between these four roles articulate through the DTOC role-based interaction model that is depicted in \autoref{fig:roles}. These roles are taken as a sufficient basis to describe the components involved in applicative tangible user interfaces specifically.

\def\legDIR{\BT{Action}}

The role of \ita{data} refers to what users sense, observe, or modify with the aid of computing resources; data justify the use or presence of such computing resources. Users perceive data through data items that serve as data presentations. In some cases, users can edit data directly through data presentations (\ie{} direct manipulation \cite{shneiderman1983direct}), and computer-generated feedback informs users about data modification. Theoretically, edited data are stored in long-term computer memory after the interaction ends. The role of data is demonstrated in various models, notations, toolkits, and paradigms, such as the instrumental interaction model (\ie{} ``domain object'') \cite{mbl2000instrumental}, the mixed interaction model (\ie{} ``task object'') \cite{coutrix2006mixed}, the ASUR notation (\ie{} ``R\textsubscript{object}'') \cite{dubois2002asur++,dubois2003asur++}, and tangible interface paradigms (\ie{} ``tokens'') \cite{holmquist1999token,shaer2004tac,ullmer2002phdthesis,ullmer2005token}.

Tangible interfaces have explored the use of physical items for several data types. Spatial data, such as terrain relief, can be manipulated directly by hand through the use of clay \cite{piper2002illclay}. Abstract data, such as the intensity of car traffic or stock exchange rates, can be conveyed by the rotation speed of Pinwheels \cite{wisneski1998pinwheels}. Likewise, the locations of mobile phone antennas can be represented by symbolic pucks on a two-dimensional city map and be moved by computing resources for optimization concerns \cite{patten2007pico}. Sometimes, data items are not editable on their own (\ie{} incapable of input), but bodied tools then allow for manipulating and modifying data.

\ita{Tools} enable users to edit data. The role of tools also appears in various models, such as the instrumental interaction model \cite{mbl2000instrumental} (\ie{} ``instrument''), the mixed interaction model \cite{coutrix2006mixed} (\ie{} ``mixed tool''), and the ASUR notation \cite{dubois2002asur++,dubois2003asur++} (\ie{} ``R\textsubscript{tool}''). For instance, in GraspDraw \cite{fitzmaurice1995bricks}, digital anchors in the form of bricks allow users to resize, rotate, and translate geometric shapes (\ie{} data editing) directly through graphical items displayed on a desk surface. In Tangible Geospace \cite{ishii1997tangible}, a device for scaling and rotating allows for browsing a campus map presentation (\ie{} data presentation editing). In Teegi \cite{frey2014teegi}, a cursor enables the indirect editing of the visualization color scheme's amplitude (\ie{} data representation editing).

\ita{Operations} provide access to the application and operating system layers. A first role of \ita{operations} is to trigger events and actions within applications and operating systems. ``Operating system operations'' cover a broad set of commands, ranging from loading or saving data \cite{ullmer2008core,ullmer2005token} to starting or closing applications \cite{ullmer2008core} and authenticating or disconnecting users \cite{wiethoff2011intuit}. ``Application operations'' allow users to alter or customize how the interactive systems behave, including the activation or parameterization of tools, the change of tool modes, the clutch or declutch of tools and data items, and the validation of actions. For instance, TUISTER \cite{butz2004tuister} enables the operation of hierarchical structures navigation across several applications. GraspDraw \cite{fitzmaurice1995bricks} features an inkwell to set the current draw color mode. In Head Prop for Neurosurgical Visualization \cite{hinckley1994passiveprops}, a thumb button and a foot pedal are used for clutching operations of the plate tool and doll's head data items, respectively. However, operations are also about exhibiting the outcomes and calculations provided by the applications and operating systems. For instance, the application of the Urban Planning workbench \cite{benjoseph2001urp,ishii2002urp,underkoffler1999urp} provides features to measure wind magnitude and the distance between buildings, which require displaying the calculated values. Another example consists of the four measurements\footnote{The four measurements of the IP Network Design Workbench \cite{kobayashi2003ip} are ``client response time,'' ``protocol response time,'' ``link usage rate,'' and ``monthly running cost.''} that are provided by the application of the IP Network Design Workbench \cite{kobayashi2003ip}, which are monitored on the vertical graph screen.

\ita{Constraints} structure interaction by guiding motion, restricting position, and organizing data, tools, or operations items. The concept initially arises as ``reference frames'' within the early tangible user interfaces framework \cite{ullmer2000emerging} and later manifests as ``constraints'' within Token + Constraints systems \cite{ullmer2002phdthesis,ullmer2005token} and the Token And Constraints paradigm \cite{shaer2004tac}. Constraints can be physical or digital.

``Physical constraints'' are imposed on bodied items before these items are sensed by computing resources. For instance, a rack orders a series of operations to process in a specific order in the Slot Machine \cite{perlman1976slotmachine}. These constraints exerted on bodied data thus indirectly fulfil a digital role by structuring mapped digital information.
Physical constraints can be visual, kinesthetic, mechanical, or dynamic:

\begin{enumerate}[label=(\alph*),ref=\alph*]
\item ``Visual constraints'' are bodiless items. They are imposed on bodied items by the users' cognition. For example, when users place pucks on a grid drawn on a whiteboard \cite{jacob2002senseboard} or when users target displayed circles with assisted pucks \cite{jobert2025agency}.
\item ``Kinesthetic constraints'' are bodied items. They are exerted on bodied items by users' cognition and sense of kinesthesia. For instance, when users place plastic cards into the rows' slots of the Slot Machine \cite{perlman1976slotmachine}.
\item ``Mechanical constraints'' are static bodied items, but they are imposed on the bodied items that are enacted or actuated by computing resources. The path that guides released balls in the Marble Answering Machine (see in \citetext{\citealp{follmer2013inform}, \citealp{ishii1997tangible}, \citealp[Ch.~1]{ullmer2022weaving}}), or a flexible curve that prevents actuated antenna representations from entering a specific area in PICO \cite{patten2007pico} are examples of such mechanical constraints.
\item ``Dynamic constraints'' are mechanical constraints that are exerted on some bodied items by dynamic bodied items (\eg{} moving or morphing constraints during the application's runtime). For example, a morphing surface that enables lifting, translating, or tilting some bodied items \cite{follmer2013inform}.
\end{enumerate}

Like in constraint-based approaches \cite{shaer2004tac,ullmer2002phdthesis,ullmer2005token}, physical constraints can eventually be associated with a set of functions that are activated when spatial concordance of some bodied items within the confining regions and that are called when sensing some physical properties of these bodied items (\eg{} position, translation, or rotation). According to the bodied component that is present, these functions can relate to data, tools, or operations.

``Digital constraints'' (also known as software constraints \cite{patten2007pico}) are exerted by computing resources on bodied items to maintain consistency with their digital state. For instance, if users disturb the equal distance between three pucks, computing resources will restore digital constraints by actuating the pucks' positions \cite{patten2007pico}.
As such, digital constraints are not physical items but the computer-generated forces exerted on bodied items.

\subsection{The Bodiness of Physical Items}

The second axis relates to the \textit{bodiness} of physical items. The three possible values comprise: static body, dynamic body, and bodiless.

\textit{Bodiless} items cannot be grasped. Such physical items encompass graphics, drawings, and sounds. Whereas the supports of graphical items are touchable (\eg{} surfaces), the graphical items themselves cannot be grasped (\ie{} held and manipulated) or sensed kinesthesically. Consequently, visual attention must be consistently maintained, preventing opportunities for eye-free operations \cite{hinckley2004techniques}. These visual items may be computer-generated graphics or handcrafted drawings on a surface. For example, the grid drawn on the Senseboard \cite{jacob2002senseboard} whiteboard serves as a bodiless constraint. The two-dimensional brain slice displayed on the distant monitor of the Head Prop for Neurosurgical Visualization \cite{hinckley1994passiveprops} is a bodiless datum item. The amplitude level displayed on the table surface of Teegi \cite{frey2014teegi} is a bodiless tool item. Likewise, audio information, which cannot be grasped is a bodiless item. Therefore, bodiless items comprise ``non-graspable'' \cite{ullmer2000emerging} and ``intangible'' \cite{ullmer2002phdthesis,ullmer2005token} representations.

Physical items with static or dynamic bodies can be grasped, as they occupy a volume in three-dimensional space---the difference between the two lies in the animation of dynamic-bodied items by computing resources. The body of \textit{static} items is inert: users can move it (\eg{} the bricks of GraspDraw \cite{fitzmaurice1995bricks}), rotate it (\eg{} the doll's head and the cutting plate in the Head Prop for Neurosurgical Visualization \cite{hinckley1994passiveprops}), and deform it (\eg{} the malleable surface of Illuminated Clay \cite{piper2002illclay} and the flexible curves of PICO \cite{patten2007pico}). However, computing resources do not affect it. Conversely, computing resources actuate the body of \textit{dynamic} items. For instance, computing resources move the antenna representations of PICO \cite{patten2007pico}, roll the cylinders of inTouch \cite{scott1997intouch}, spin the blades of Pinwheels \cite{wisneski1998pinwheels}, and alter the shape of Relief \cite{leithinger2010relief,leithinger2011relief}.

Consequently, this taxonomy establishes a conceptual space consisting of twelve categories that account for components' roles and their items' bodiness. The remainder of this section builds upon this taxonomy to classify tangible interfaces into four classes.

\subsection{Four Role-Centered Bodiness Classes}\label{sec-classification}

This section suggests classifying applicative tangible interfaces based on the bodiness of data items:

\begin{itemize}

\item \textit{Class~I specimens} exclusively rely on bodied data, using either static or dynamic bodied data items without bodiless data items.

\item \textit{Class~II specimens} rely on a combination of bodiless data items with either static or dynamic bodied data items.

\item \textit{Class~III specimens} are based on the combination of bodied tools with bodiless data (\ie{} bodied data are absent).

\item \textit{Class~IV specimens} only provide bodied operations (\ie{} no data) that are not restricted to a single application and can control the same shared operations across applications.

\end{itemize}

\autoref{tab::classes} presents the occurrence patterns of item kinds in each class. In addition, we can rediscover some of the major research inspirations in tangible interfaces in these four classes:
\begin{itemize}
\item Pure (or ``full'' \cite{fishkin2004taxonomy}) tangible user interfaces (\ie{} the most tangible ones, in regard to the ``tangible bits'' vision \cite{ishii1997tangible} and the tangibility spectrum \cite{fishkin2004taxonomy}) belong to Class~I;
\item Passive real-world props \cite{hinckley1994passiveprops} belong to Class~II;
\item Graspable user interfaces \cite{fitzmaurice1995bricks} belong to Class~III;
\item Core tangibles \cite{ullmer2008core} belong to Class~IV.
\end{itemize}
  
The following section analyzes the literature by using the taxonomy to describe the items from a sample of applications, which are then classified into the four classes.

\def\c{\centering}

\begin{table}
{\centering\small
  \begin{tabular}{lllp{8mm}p{8mm}p{8mm}p{8mm}}
\midrule
\multicolumn{2}{l}{\textbf{\ita{Physical Item Kind}}} & & \multicolumn{4}{l}{\textbf{\ita{Bodiness Class}}} \tabularnewline
\cmidrule{1-2}\cmidrule{4-7}
\ita{Role} & \ita{Bodiness} && \ita{Class~I} & \ita{Class~II} & \ita{Class~III} & \ita{Class~IV} \tabularnewline    
\midrule

Data        & Bodied    &&\c +      &\c +      &\c 0      &\c 0      \tabularnewline\vspace{2pt}
            & Bodiless  &&\c 0      &\c +      &\c +      &\c 0      \tabularnewline
Tools       & Bodied    &&\c $\ast$ &\c $\ast$ &\c +      &\c 0      \tabularnewline\vspace{2pt}
            & Bodiless  &&\c $\ast$ &\c $\ast$ &\c $\ast$ &\c 0      \tabularnewline
Operations  & Bodied    &&\c $\ast$ &\c $\ast$ &\c $\ast$ &\c +      \tabularnewline\vspace{2pt}
            & Bodiless  &&\c $\ast$ &\c $\ast$ &\c $\ast$ &\c $\ast$ \tabularnewline
Constraints & Bodied    &&\c $\ast$ &\c $\ast$ &\c $\ast$ &\c $\ast$ \tabularnewline
            & Bodiless  &&\c $\ast$ &\c $\ast$ &\c $\ast$ &\c $\ast$ \tabularnewline

\midrule
  \end{tabular}\\
  \scriptsize{
    \ita{Note.} ``+'' = some items (one or more). ``$\ast$'' = none or some items.
  }
}
\caption{Definition of four bodiness classes according to the physical item involved in the four components' roles.}~\label{tab::classes}
\vspace{-9pt}
\end{table}

\section{Analyzing Applicative Tangible Interfaces}\label{sec-analysis}

This section selects a collection of applicative tangible interfaces from the literature that deal with various application domains, and describes the items used in this collection through the taxonomy and sorts the applications into the four classes. Then, it scrutinizes the evolution of the field and the design of applicative tangible interfaces by examining the representative sample of specimens.

\subsection{Selection Process of the Collection}\label{sec-collection}

The current collection is intended to be merely a partial summary of hundreds of tangible interface specimens that researchers produced in the past few decades. Specimen selection is less strict than in systematic literature reviews, which hinders the performance of precise statistics. The goal of the selection is to encompass a variety of emblematic tangible interface specimens and offer a representative overview alongside some insights into the field. Restricting the selection to a specific date range may have led to the exclusion of certain tangible interface kinds, as research interests fluctuate over the years. Thus, the entire research period was taken into account, from premises to recent work.
Six relevant sources were browsed:

\begin{itemize}
 
\item The project page\footnote{The project page of the Tangible Media Group displays a chronological list of specimen names and corresponding thumbnail images: \url{https://tangible.media.mit.edu/projects/} (last accessed on August 31, 2023).} of the MIT Media Lab's Tangible Media Group (TMG), which is recognized as a prominent and prolific group in the exploration of tangible user interfaces. The final collection includes specimens of this provenance in \purple{43\%} of the cases.
  
\item The sessions with ``tangible,'' ``embodied,'' ``physical,'' or related terms in their title from \purple{eighteen} ACM CHI conference proceedings spanning from \purple{2005 to 2022} venues\footnote{As of August 31, 2023, the titles of ACM CHI conference sessions are available on the ACM Digital Library website, \url{https://dl.acm.org/}, from the year 2005 onward. No dedicated session was found for the year 2023.};
  
\item The \purple{seventeen} ACM TEI conference proceedings, from \purple{2007 to 2023} venues;

\item Three states of the art, from one book devoted to the field in 2022 \cite{ullmer2022weaving}, one major review of the field in 2010 \cite{shaer2010past}, and one literature study that focused on applicative tangible interfaces in 2009 \cite{riviere2009phd}. 

\end{itemize}

The sources were screened to retain work dealing with applications. Work solely demonstrating a new technology, interaction technique, or tangible interface kind was excluded because describing them with the DTOC role-based interaction model makes no sense. Entries were filtered based on their name, thumbnail image, or publication title. Therefore, entries that provided a sneak peek of an application in their name, thumbnail image, or publication title were given precedence. Publications and demonstration videos of eligible work were downloaded and filtered to ensure adequate descriptions of applications, user interfaces, and interactions. Only applications with new application domains were included in the collection, unless they had different implementations from other ones. \autoref{tab::selection} provides a summary of the \purple{six} sources' inclusion in the collection.

\begin{table}
{\centering\small
  \begin{tabular}{
  l@{\hspace{5pt}}l@{\hspace{4pt}}m{0mm}
  @{\hspace{4pt}}m{6mm}@{\hspace{4pt}}m{6.5mm}@{\hspace{4pt}}m{7mm}@{\hspace{4pt}}m{8mm}@{\hspace{4pt}}m{8mm}@{\hspace{4pt}}m{6mm}@{\hspace{4pt}}m{6mm}
  }
\midrule
\multicolumn{2}{l}{\itb{Source}} && \multicolumn{7}{l}{\itb{Specimens' Inclusion}}  \tabularnewline
\cmidrule{1-2}\cmidrule{4-10}
\ita{Name} & \ita{Period}  && \ita{Once} & \ita{Twice} & \ita{Thrice} & \ita{Quarce} & \ita{Quince} & \ita{Sence}   & \ita{Total} \tabularnewline
\midrule
TMG                                                     & 1990-2023  &&  \raL 1& \raL 2& \raL 3& \raL 4& \raL 5 & \raL 0 & \raL 15\tabularnewline
CHI                                                     & 2005-2022  &&  \raL 2& \raL 2& \raL 4& \raL 6& \raL 5 & \raL 0 & \raL 19\tabularnewline
TEI                                                     & 2007-2023  &&  \raL 1& \raL 2& \raL 3& \raL 1& \raL 0 & \raL 0 & \raL 7 \tabularnewline
\cite{ullmer2022weaving}                                & 1976-2022  &&  \raL 1& \raL 5& \raL 0& \raL 5& \raL 5 & \raL 0 & \raL 16\tabularnewline 
\cite{shaer2010past}                                    & 1976-2010  &&  \raL 0& \raL 5& \raL 7& \raL 5& \raL 5 & \raL 0 & \raL 22\tabularnewline 
\cite{riviere2009phd}                                   & 1991-2009  &&  \raL 3& \raL 0& \raL 4& \raL 7& \raL 5 & \raL 0 & \raL 19\tabularnewline
\midrule
{All}                                                   & 1976-2023  &&  \raL 8& \raL 8& \raL 7& \raL 7& \raL 5 & \raL 0   & \raL 35 \tabularnewline
&            &&  \raL 23\% & \raL 23\% & \raL 20\% & \raL 20\% & \raL 14\% & \raL 0\% & \tabularnewline
\midrule
  \end{tabular}}
\caption{Composition of the collection of \numapps{} specimens from six sources, showing overlaps between sources.}~\label{tab::selection}
\vspace{-9pt}
\end{table}

\subsection{Applications' Labeling and Classification}\label{sec-labeling}

Inventorying and labeling of the physical items were carried out by reading their descriptions in publications and watching any available supplementary demonstration videos.
Tables~\ref{tab::illustration1}, \ref{tab::illustration2}, and~\ref{tab::illustration3} report the resulting labeling of \numreps{} physical items from the \numapps{} applications of the collection. The domains and bodiness classes are listed in \autoref{tab::by-classes}: the collection comprises \purple{4} applications of Class~I, \purple{21} of Class~II, \purple{5} of Class~III, and \purple{5} of Class~IV.

\def\captionI{The physical items of \purple{thirtheen} applicative tangible interfaces (ordered by ascending years, from \purple{1976} to \purple{2001}).}

\def\captionII{The physical items of \purple{fourteen} applicative tangible interfaces (ordered by ascending years, from \purple{2002} to \purple{2010}).}

\def\captionIII{The physical items of \purple{eigth} applicative tangible interfaces (ordered by ascending years, from \purple{2011} to \purple{2022}).}

\def\legend{\vspace{-2pt}\ita{Note.} ``-'' = empty. Role: ``D'' = Data, ``T'' = Tool, ``O'' = Operation, ``C'' = Constraint. Bodiness: ``S'' = Static, ``D'' = Dynamic, ``L'' = Bodiless.} 

\begin{table*}
{\centering\footnotesize
  \begin{tabular}{rlllllp{0cm}llllllll}

    \midrule

\multicolumn{4}{l}{\textbf{\textit{Tangible Interface}}}
    & & \multicolumn{10}{l}{\textbf{\textit{Physical Items}}} \tabularnewline

    \cmidrule{1-4}\cmidrule{6-15}
    
\multicolumn{4}{l}{\textbf{\textit{}}}
    & &
    & & \multicolumn{4}{l}{\textbf{\textit{Role}}}
    & & \multicolumn{3}{l}{\textbf{\textit{Bodiness}}}
    \tabularnewline
    
    \cmidrule{8-11}\cmidrule{13-15}

\textit{\#}
    & \textit{Name}
    & \textit{Year}
    & \textit{References}
    & & \textit{Name or Description}
    & & \textit{D}
    & \textit{T}
    & \textit{O}
    & \textit{C}
    & & \textit{S}
    & \textit{D}
    & \textit{L}
    \tabularnewline
    
    \midrule
    
    1.&Slot Machine&1976&\cite{perlman1976slotmachine}&&Rows (procedure)&&\o{}&\o{}&\o{}&\xc{}&&\xs{}&\o{}&\o{}\tabularnewline
&&&&&Buttons (on each row)&&\o{}&\o{}&\xo{}&\o{}&&\xs{}&\o{}&\o{}\tabularnewline
&&&&&Plastic cards (commands)&&\o{}&\o{}&\xo{}&\o{}&&\xs{}&\o{}&\o{}\tabularnewline
&&&&&Display triangle (turtle)&&\xd{}&\o{}&\o{}&\o{}&&\o{}&\o{}&\xl{}\tabularnewline
&&&&&Circular robot (turtle)&&\xd{}&\o{}&\o{}&\o{}&&\xs{}&\o{}&\o{}\vspace{2pt}\tabularnewline

2.&CAAD 3D Modelling System&1979&\cite{aish1979caad,aish1984caad}&&Building Blocks&&\xd{}&\o{}&\o{}&\o{}&&\xs{}&\o{}&\o{}\tabularnewline
&&&&&Design geometry (perspective view)&&\xd{}&\o{}&\o{}&\o{}&&\o{}&\o{}&\xl{}\tabularnewline
&&&&&Evaluation measures (Isoplots)&&\xd{}&\o{}&\o{}&\o{}&&\o{}&\o{}&\xl{}\vspace{2pt}\tabularnewline

3.&Self-Builder Model (Segal Model)&1980&\cite{frazer1982three,frazer1980intelligent} cited from \cite{sutphen2000reviving}&&Panels&&\xd{}&\o{}&\o{}&\o{}&&\xs{}&\o{}&\o{}\tabularnewline
&&&&&Board&&\o{}&\o{}&\o{}&\xc{}&&\xs{}&\o{}&\o{}\tabularnewline
&&&&&Wireframe rendering&&\xd{}&\o{}&\o{}&\o{}&&\o{}&\o{}&\xl{}\tabularnewline
&&&&&Feedback tool (house area, cost)&&\xd{}&\o{}&\o{}&\o{}&&\o{}&\o{}&\xl{}\vspace{2pt}\tabularnewline

4.&Marble Answering Machine&1992&See in \citetext{\citealp{follmer2013inform}, \citealp{ishii1997tangible}, \citealp[Ch.~1]{ullmer2022weaving}}&&Machine's path (queue)&&\o{}&\o{}&\o{}&\xc{}&&\xs{}&\o{}&\o{}\tabularnewline
&&&&&Machine's indentation (play slot)&&\o{}&\o{}&\xo{}&\o{}&&\xs{}&\o{}&\o{}\tabularnewline
&&&&&Marble (message)&&\xd{}&\o{}&\o{}&\o{}&&\o{}&\xy{}&\o{}\tabularnewline
&&&&&Message&&\xd{}&\o{}&\o{}&\o{}&&\o{}&\o{}&\xl{}\vspace{2pt}\tabularnewline

5.&Head Prop&1994&\cite{hinckley1994passiveprops}&&Doll's head (brain)&&\xd{}&\o{}&\o{}&\o{}&&\xs{}&\o{}&\o{}\tabularnewline
&&&&&Plate (slicing)&&\o{}&\xt{}&\o{}&\o{}&&\xs{}&\o{}&\o{}\tabularnewline
&&&&&Plate thumb button (clutch)&&\o{}&\o{}&\xo{}&\o{}&&\xs{}&\o{}&\o{}\tabularnewline
&&&&&Foot pedal (clutch)&&\o{}&\o{}&\xo{}&\o{}&&\xs{}&\o{}&\o{}\tabularnewline
&&&&&Brain 2D slice&&\xd{}&\o{}&\o{}&\o{}&&\o{}&\o{}&\xl{}\vspace{2pt}\tabularnewline

6.&GraspDraw&1995&\cite{fitzmaurice1995bricks}&&Bricks&&\o{}&\xt{}&\o{}&\o{}&&\xs{}&\o{}&\o{}\tabularnewline
&&&&&2D shapes&&\xd{}&\o{}&\o{}&\o{}&&\o{}&\o{}&\xl{}\tabularnewline
&&&&&Inkwell&&\o{}&\o{}&\xo{}&\o{}&&\xs{}&\o{}&\o{}\tabularnewline
&&&&&Functions' tray (select, delete, shapes)&&\o{}&\o{}&\xo{}&\o{}&&\xs{}&\o{}&\o{}\tabularnewline
&&&&&ActiveDesk surface&&\o{}&\o{}&\o{}&\xc{}&&\xs{}&\o{}&\o{}\vspace{2pt}\tabularnewline

7.&Tangible Geospace&1997&\cite{ishii1997tangible}&&Phicons&&\xd{}&\o{}&\o{}&\o{}&&\xs{}&\o{}&\o{}\tabularnewline
&&&&&Campus Map&&\xd{}&\o{}&\o{}&\o{}&&\o{}&\o{}&\xl{}\tabularnewline
&&&&&Overlay view (passive lens)&&\o{}&\o{}&\xo{}&\o{}&&\xs{}&\o{}&\o{}\tabularnewline
&&&&&3D view (active lens)&&\xd{}&\o{}&\o{}&\o{}&&\o{}&\o{}&\xl{}\tabularnewline
&&&&&Scaling and Rotating Device&&\o{}&\xt{}&\o{}&\o{}&&\xs{}&\o{}&\o{}\tabularnewline
&&&&&Desk's surface&&\o{}&\o{}&\o{}&\xc{}&&\xs{}&\o{}&\o{}\vspace{2pt}\tabularnewline

8.&BUILD-IT&1997&\cite{rauterberg1997buildit,rauterberg1998buildit,fjeld1999camera}&&Bricks&&\o{}&\xt{}&\o{}&\o{}&&\xs{}&\o{}&\o{}\tabularnewline
&&&&&Plan view&&\xd{}&\o{}&\o{}&\o{}&&\o{}&\o{}&\xl{}\tabularnewline
&&&&&Objects (robots, tables\ldots)&&\xd{}&\o{}&\o{}&\o{}&&\o{}&\o{}&\xl{}\tabularnewline
&&&&&Virtual cameras&&\o{}&\o{}&\xo{}&\o{}&&\o{}&\o{}&\xl{}\tabularnewline
&&&&&EyeCatchers&&\o{}&\o{}&\xo{}&\o{}&&\o{}&\o{}&\xl{}\tabularnewline
&&&&&Table surface&&\o{}&\o{}&\o{}&\xc{}&&\o{}&\o{}&\o{}\tabularnewline
&&&&&3D view&&\xd{}&\o{}&\o{}&\o{}&&\xs{}&\o{}&\xl{}\vspace{2pt}\tabularnewline

9.&Pinwheels&1998&\cite{wisneski1998pinwheels}&&Pinwheels&&\xd{}&\o{}&\o{}&\o{}&&\o{}&\xy{}&\o{}\vspace{2pt}\tabularnewline

10.&mediaBlocks&1998&\cite{ullmer1999mediablocks,ullmer1998mediablocks}&&mediaBlock (storage)&&\o{}&\o{}&\xo{}&\o{}&&\xs{}&\o{}&\o{}\tabularnewline
&&&&&Slots&&\o{}&\o{}&\o{}&\xc{}&&\xs{}&\o{}&\o{}\vspace{2pt}\tabularnewline

11.&musicBottles&1999&\cite{ishii1999musicbottles,ishii2001musicbottles}&&Music&&\xd{}&\o{}&\o{}&\o{}&&\o{}&\o{}&\xl{}\tabularnewline
&&&&&Bottle (music)&&\xd{}&\o{}&\o{}&\o{}&&\xs{}&\o{}&\o{}\tabularnewline
&&&&&Cork&&\o{}&\o{}&\xo{}&\o{}&&\xs{}&\o{}&\o{}\tabularnewline
&&&&&Triangular table&&\o{}&\o{}&\o{}&\xc{}&&\xs{}&\o{}&\o{}\tabularnewline
&&&&&Central ``stage'' area&&\o{}&\o{}&\o{}&\xc{}&&\o{}&\o{}&\xl{}\vspace{2pt}\tabularnewline

12.&Urp (Urban Planning Workbench)&1999&\cite{benjoseph2001urp,ishii2002urp,underkoffler1999urp}&&Architectural Model&&\xd{}&\o{}&\o{}&\o{}&&\xs{}&\o{}&\o{}\tabularnewline
&&&&&Road-object (strips)&&\xd{}&\o{}&\o{}&\o{}&&\xs{}&\o{}&\o{}\tabularnewline
&&&&&Material-transformation-object (wand)&&\o{}&\xt{}&\o{}&\o{}&&\xs{}&\o{}&\o{}\tabularnewline
&&&&&Video-camera-object&&\o{}&\xt{}&\o{}&\o{}&&\xs{}&\o{}&\o{}\tabularnewline
&&&&&Clock-object&&\o{}&\xt{}&\o{}&\o{}&&\xs{}&\o{}&\o{}\tabularnewline
&&&&&Wind-generating tool&&\o{}&\xt{}&\o{}&\o{}&&\xs{}&\o{}&\o{}\tabularnewline
&&&&&Anemometer-object (arrow)&&\o{}&\o{}&\xo{}&\o{}&&\xs{}&\o{}&\o{}\tabularnewline
&&&&&Wind magnitude (number)&&\o{}&\o{}&\xo{}&\o{}&&\o{}&\o{}&\xl{}\tabularnewline
&&&&&Distance-measuring-object&&\o{}&\o{}&\xo{}&\o{}&&\xs{}&\o{}&\o{}\tabularnewline
&&&&&Distance (number)&&\o{}&\o{}&\xo{}&\o{}&&\o{}&\o{}&\xl{}\tabularnewline
&&&&&Shadows/Reflections&&\xd{}&\o{}&\o{}&\o{}&&\o{}&\o{}&\xl{}\tabularnewline
&&&&&Airflow grid&&\xd{}&\o{}&\o{}&\o{}&&\o{}&\o{}&\xl{}\tabularnewline
&&&&&Workbench&&\o{}&\o{}&\o{}&\xc{}&&\xs{}&\o{}&\o{}\vspace{2pt}\tabularnewline

13.&Senseboard&2001&\cite{jacob2002senseboard}&&Grid&&\o{}&\o{}&\o{}&\xc{}&&\o{}&\o{}&\xl{}\tabularnewline
&&&&&Rectangular pucks&&\xd{}&\o{}&\o{}&\o{}&&\xs{}&\o{}&\o{}\tabularnewline
&&&&&View detail puck&&\o{}&\xt{}&\o{}&\o{}&&\xs{}&\o{}&\o{}\tabularnewline
&&&&&Arrow puck&&\o{}&\xt{}&\o{}&\o{}&&\xs{}&\o{}&\o{}\tabularnewline
&&&&&Values&&\xd{}&\o{}&\o{}&\o{}&&\o{}&\o{}&\xl{}\tabularnewline
&&&&&Board&&\o{}&\o{}&\o{}&\xc{}&&\xs{}&\o{}&\o{}\vspace{2pt}\tabularnewline

14.&Illuminating Clay&2002&\cite{ishii2004illclay,piper2002illclay}&&Clay Model&&\xd{}&\o{}&\o{}&\o{}&&\xs{}&\o{}&\o{}\tabularnewline
&&&&&Rotative platform&&\o{}&\o{}&\o{}&\xc{}&&\xs{}&\o{}&\o{}\tabularnewline
 
    \midrule
    
  \end{tabular}\\
      \scriptsize{\legend}
}
\caption{\captionI}~\label{tab::illustration1}
\vspace{-9pt}
\end{table*}

\begin{table*}
{\centering\footnotesize
  \begin{tabular}{rllllllllllllll}

    \midrule

\multicolumn{4}{l}{\textbf{\textit{Tangible Interface}}}
    & & \multicolumn{10}{l}{\textbf{\textit{Physical Items}}} \tabularnewline

    \cmidrule{1-4}\cmidrule{6-15}
    
\multicolumn{4}{l}{\textbf{\textit{}}}
    & &
    & & \multicolumn{4}{l}{\textbf{\textit{Role}}}
    & & \multicolumn{3}{l}{\textbf{\textit{Bodiness}}}
    \tabularnewline
    
    \cmidrule{8-11}\cmidrule{13-15}

\textit{\#}
    & \textit{Name}
    & \textit{Year}
    & \textit{References}
    & & \textit{Name or Description}
    & & \textit{D}
    & \textit{T}
    & \textit{O}
    & \textit{C}
    & & \textit{S}
    & \textit{D}
    & \textit{L}
    \tabularnewline
    
    \midrule
    
    &&&&&Crosshairs&&\o{}&\xt{}&\o{}&\o{}&&\o{}&\o{}&\xl{}\tabularnewline
&&&&&Cross Sections&&\xd{}&\o{}&\o{}&\o{}&&\o{}&\o{}&\xl{}\tabularnewline
&&&&&Analysis Function Thumbnails&&\xd{}&\o{}&\o{}&\o{}&&\o{}&\o{}&\xl{}\tabularnewline
&&&&&3-D perspective view&&\xd{}&\o{}&\o{}&\o{}&&\o{}&\o{}&\xl{}\vspace{2pt}\tabularnewline

15.&Audiopad&2002&\cite{patten2002audiopad,patten2006audiopad}&&Sounds&&\xd{}&\o{}&\o{}&\o{}&&&\o{}&\xl{}\tabularnewline
&&&&&Pucks (audio tracks)&&\xd{}&\o{}&\o{}&\o{}&&\xs{}&\o{}&\o{}\tabularnewline
&&&&&Star-shape puck (sound selector)&&\o{}&\xt{}&\o{}&\o{}&&\xs{}&\o{}&\o{}\tabularnewline
&&&&&Hierarchical menu&&\o{}&\o{}&\xo{}&\o{}&&\o{}&\o{}&\xl{}\tabularnewline
&&&&&SenseTable surface&&\o{}&\o{}&\o{}&\xc{}&&\xs{}&\o{}&\o{}\vspace{2pt}\tabularnewline

16.&reacTable&2003&\cite{jorda2007reactable}&&Music&&\xd{}&\o{}&\o{}&\o{}&&\o{}&\o{}&\xl{}\tabularnewline
&&&&&Point (audio output)&&\xd{}&\o{}&\o{}&\o{}&&\o{}&\o{}&\xl{}\tabularnewline
&&&&&Square puck (audio source)&&\xd{}&\o{}&\o{}&\o{}&&\xs{}&\o{}&\o{}\tabularnewline
&&&&&Rounded square puck (filter)&&\o{}&\xt{}&\o{}&\o{}&&\xs{}&\o{}&\o{}\tabularnewline
&&&&&Round puck (controller)&&\o{}&\xt{}&\o{}&\o{}&&\xs{}&\o{}&\o{}\tabularnewline
&&&&&Decagon puck (control filter)&&\o{}&\xt{}&\o{}&\o{}&&\xs{}&\o{}&\o{}\tabularnewline
&&&&&Pentagon puck (audio mixer)&&\o{}&\xt{}&\o{}&\o{}&&\xs{}&\o{}&\o{}\tabularnewline
&&&&&Lines (audio flow)&&\xd{}&\o{}&\o{}&\o{}&&\o{}&\o{}&\xl{}\tabularnewline
&&&&&Table surface&&\o{}&\o{}&\o{}&\xc{}&&\xs{}&\o{}&\o{}\vspace{2pt}\tabularnewline

17.&IP Network Design Workbench&2003&\cite{kobayashi2003ip}&&Pucks&&\o{}&\xt{}&\o{}&\o{}&&\xs{}&\o{}&\o{}\tabularnewline
&&&&&Nodes&&\xd{}&\o{}&\o{}&\o{}&&\o{}&\o{}&\xl{}\tabularnewline
&&&&&Links&&\xd{}&\o{}&\o{}&\o{}&&\o{}&\o{}&\xl{}\tabularnewline
&&&&&Nodes menu&&\o{}&\o{}&\xo{}&\o{}&&\o{}&\o{}&\xl{}\tabularnewline
&&&&&Parameter puck (with button)&&\o{}&\xt{}&\o{}&\o{}&&\xs{}&\o{}&\o{}\tabularnewline
&&&&&Link bandwidth&&\o{}&\xt{}&\o{}&\o{}&&\o{}&\o{}&\xl{}\tabularnewline
&&&&&Router service priority&&\o{}&\xt{}&\o{}&\o{}&&\o{}&\o{}&\xl{}\tabularnewline
&&&&&Number of client users&&\o{}&\xt{}&\o{}&\o{}&&\o{}&\o{}&\xl{}\tabularnewline
&&&&&Server performance&&\o{}&\xt{}&\o{}&\o{}&&\o{}&\o{}&\xl{}\tabularnewline
&&&&&Table surface&&\o{}&\o{}&\o{}&\xc{}&&\xs{}&\o{}&\o{}\tabularnewline
&&&&&Measurement graphs&&\o{}&\o{}&\xo{}&\o{}&&\o{}&\o{}&\xl{}\vspace{2pt}\tabularnewline

18.&Query Shapes&2004&\cite{ichida2004retrieval}&&Cubes (ActiveCubes)&&\xd{}&\o{}&\o{}&\o{}&&\xs{}&\o{}&\o{}\tabularnewline
&&&&&Voxel data representation&&\xd{}&\o{}&\o{}&\o{}&&\o{}&\o{}&\xl{}\tabularnewline
&&&&&3D shape models&&\xd{}&\o{}&\o{}&\o{}&&\o{}&\o{}&\xl{}\vspace{2pt}\tabularnewline

19.&TUISTER&2004&\cite{butz2004tuister}&&TUISTER&&\o{}&\o{}&\xo{}&\o{}&&\xs{}&\o{}&\o{}\vspace{2pt}\tabularnewline

20.&I/O Brush&2004&\cite{ryokai2004iobrush,ryokai2007iobrush}&&Brush&&\o{}&\xt{}&\o{}&\o{}&&\xs{}&\o{}&\o{}\tabularnewline
&&&&&World (palette)&&\o{}&\o{}&\xo{}&\o{}&&\xs{}&\o{}&\o{}\tabularnewline
&&&&&Canvas&&\xd{}&\o{}&\o{}&\o{}&&\o{}&\o{}&\xl{}\vspace{2pt}\tabularnewline

21.&PICO&2005&\cite{patten2007pico}&&Pucks (antenas)&&\xd{}&\o{}&\o{}&\o{}&&\o{}&\xy{}&\o{}\tabularnewline
&&&&&Map (city)&&\xd{}&\o{}&\o{}&\o{}&&\o{}&\o{}&\xl{}\tabularnewline
&&&&&Flexible curve&&\o{}&\o{}&\o{}&\xc{}&&\xs{}&\o{}&\o{}\tabularnewline
&&&&&Rubber band&&\o{}&\o{}&\o{}&\xc{}&&\xs{}&\o{}&\o{}\tabularnewline
&&&&&Collar&&\o{}&\o{}&\o{}&\xc{}&&\xs{}&\o{}&\o{}\tabularnewline
&&&&&Teflon / Sandpaper&&\o{}&\o{}&\o{}&\xc{}&&\xs{}&\o{}&\o{}\tabularnewline
&&&&&Table surface&&\o{}&\o{}&\o{}&\xc{}&&\xs{}&\o{}&\o{}\vspace{2pt}\tabularnewline

22.&Augmented Chemistry&2007&\cite{fjeld2007chemistry}&&Booklet&&\o{}&\o{}&\xo{}&\o{}&&\xs{}&\o{}&\o{}\tabularnewline
&&&&&Gripper&&\o{}&\xt{}&\o{}&\o{}&&\xs{}&\o{}&\o{}\tabularnewline
&&&&&Cube&&\xd{}&\o{}&\o{}&\o{}&&\xs{}&\o{}&\o{}\tabularnewline
&&&&&Augmented 3D view&&\xd{}&\o{}&\o{}&\o{}&&\o{}&\o{}&\xl{}\vspace{2pt}\tabularnewline

23.&Nimio&2007&\cite{brewer2007nimio}&&Pyramid Nimios (color-changing)&&\xd{}&\o{}&\o{}&\o{}&&\xs{}&\o{}&\o{}\tabularnewline
&&&&&Cube Nimios (color-changing)&&\xd{}&\o{}&\o{}&\o{}&&\xs{}&\o{}&\o{}\tabularnewline
&&&&&Dome Nimios (color-changing)&&\xd{}&\o{}&\o{}&\o{}&&\o{}&\o{}&\o{}\tabularnewline
&&&&&Cylinder Nimios (color-changing)&&\xd{}&\o{}&\o{}&\o{}&&\xs{}&\o{}&\o{}\vspace{2pt}\tabularnewline

24.&ArcheoTUI&2007&\cite{reuter2007archeotui}&&Props&&\xd{}&\o{}&\o{}&\o{}&&\xs{}&\o{}&\o{}\tabularnewline
&&&&&3D fragments&&\xd{}&\o{}&\o{}&\o{}&&\o{}&\o{}&\xl{}\tabularnewline
&&&&&Foot pedals (clutch)&&\o{}&\o{}&\xo{}&\o{}&&\xs{}&\o{}&\o{}\vspace{2pt}\tabularnewline

25.&GeoTUI&2008&\cite{couture2008geotui}&&Map (cube top view)&&\xd{}&\o{}&\o{}&\o{}&&\o{}&\o{}&\xl{}\tabularnewline
&&&&&Cutting line&&\o{}&\xt{}&\o{}&\o{}&&\o{}&\o{}&\xl{}\tabularnewline
&&&&&Two-puck prop&&\o{}&\xt{}&\o{}&\o{}&&\xs{}&\o{}&\o{}\tabularnewline
&&&&&Ruler prop&&\o{}&\xt{}&\o{}&\o{}&&\xs{}&\o{}&\o{}\tabularnewline
&&&&&Button box&&\o{}&\o{}&\xo{}&\o{}&&\xs{}&\o{}&\o{}\tabularnewline
&&&&&Tabletop&&\o{}&\o{}&\o{}&\xc{}&&\xs{}&\o{}&\o{}\vspace{2pt}\tabularnewline

26.&Slurp&2008&\cite{zigelbaum2008slurp}&&Eyedropper&&\o{}&\o{}&\xo{}&\o{}&&\xs{}&\o{}&\o{}\vspace{2pt}\tabularnewline

27.&Relief&2010&\cite{leithinger2010relief,leithinger2011relief}&&2.5D shape display (terrain)&&\xd{}&\o{}&\o{}&\o{}&&\o{}&\xy{}&\o{}\tabularnewline
&&&&&Topographical map&&\xd{}&\o{}&\o{}&\o{}&&\o{}&\o{}&\xl{}\vspace{2pt}\tabularnewline

28.&Bidirectional mixiTUI&2011&\cite{pedersen2009mixitui,pedersen2011tangiblebots}&&Sounds (Music)&&\xd{}&\o{}&\o{}&\o{}&&\o{}&\o{}&\xl{}\tabularnewline
&&&&&Tracks (vibrating lines)&&\xd{}&\o{}&\o{}&\o{}&&\o{}&\o{}&\xl{}\tabularnewline
 
    \midrule
    
  \end{tabular}\\
    \scriptsize{\legend}
}
\caption{\captionII}~\label{tab::illustration2}
\vspace{-9pt}
\end{table*}

\begin{table*}
{\centering\footnotesize
  \begin{tabular}{rllllllllllllll}

    \midrule

\multicolumn{4}{l}{\textbf{\textit{Tangible Interface}}}
    & & \multicolumn{10}{l}{\textbf{\textit{Physical Items}}} \tabularnewline

    \cmidrule{1-4}\cmidrule{6-15}
    
\multicolumn{4}{l}{\textbf{\textit{}}}
    & &
    & & \multicolumn{4}{l}{\textbf{\textit{Role}}}
    & & \multicolumn{3}{l}{\textbf{\textit{Bodiness}}}
    \tabularnewline
    
    \cmidrule{8-11}\cmidrule{13-15}

\textit{\#}
    & \textit{Name}
    & \textit{Year}
    & \textit{References}
    & & \textit{Name or Description}
    & & \textit{D}
    & \textit{T}
    & \textit{O}
    & \textit{C}
    & & \textit{S}
    & \textit{D}
    & \textit{L}
    \tabularnewline
    
    \midrule
    
    &&&&&Active loop tokens (Precomposed samples)&&\xd{}&\o{}&\o{}&\o{}&&\o{}&\xy{}&\o{}\tabularnewline
&&&&&Loop tokens graphical feedback&&\xd{}&\o{}&\o{}&\o{}&&\o{}&\o{}&\xl{}\tabularnewline
&&&&&Active effect tokens&&\o{}&\xt{}&\o{}&\o{}&&\o{}&\xy{}&\o{}\tabularnewline
&&&&&Effect tokens graphical feedback&&\o{}&\xt{}&\o{}&\o{}&&\o{}&\o{}&\xl{}\tabularnewline
&&&&&Control tokens (Session, Stop)&&\o{}&\o{}&\xo{}&\o{}&&\xs{}&\o{}&\o{}\tabularnewline
&&&&&Colored tabs (sessions selectors)&&\o{}&\o{}&\xo{}&\o{}&&\o{}&\o{}&\xl{}\vspace{2pt}\tabularnewline

29.&Teegi&2014&\cite{frey2014teegi}&&Teegi character (user's activity)&&\xd{}&\o{}&\o{}&\o{}&&\xs{}&\o{}&\o{}\tabularnewline
&&&&&Brain model (user's activity)&&\xd{}&\o{}&\o{}&\o{}&&\xs{}&\o{}&\o{}\tabularnewline
&&&&&Filter area&&\o{}&\xt{}&\o{}&\o{}&&\o{}&\o{}&\xl{}\tabularnewline
&&&&&Mini-Teegis (filters)&&\o{}&\xt{}&\o{}&\o{}&&\xs{}&\o{}&\o{}\tabularnewline
&&&&&EEG Raw Signals&&\xd{}&\o{}&\o{}&\o{}&&\o{}&\o{}&\xl{}\tabularnewline
&&&&&Color map amplitude&&\o{}&\xt{}&\o{}&\o{}&&\o{}&\o{}&\xl{}\tabularnewline
&&&&&Color map cursor &&\o{}&\xt{}&\o{}&\o{}&&\xs{}&\o{}&\o{}\tabularnewline
&&&&&Table surface&&\o{}&\o{}&\o{}&\xc{}&&\xs{}&\o{}&\o{}\vspace{2pt}\tabularnewline

30.&SoundFORMS&2016&\cite{colter2016soundforms}&&Trigger pins&&\o{}&\o{}&\xo{}&\o{}&&\xs{}&\o{}&\o{}\tabularnewline
&&&&&Soundwave pins&&\xd{}&\o{}&\o{}&\o{}&&\o{}&\xy{}&\o{}\tabularnewline
&&&&&Sounds&&\xd{}&\o{}&\o{}&\o{}&&\o{}&\o{}&\xl{}\vspace{2pt}\tabularnewline

31.&CairnFORM&2019&\cite{daniel2019cairnform}&&Ring chart&&\xd{}&\o{}&\o{}&\o{}&&\o{}&\xy{}&\o{}\vspace{2pt}\tabularnewline

32.&reSpire&2019&\cite{choi2019respire}&&Shape-changing fabric&&\xd{}&\o{}&\o{}&\o{}&&\o{}&\xy{}&\o{}\vspace{2pt}\tabularnewline

33.&Embodied Axes&2020&\cite{cordeil2020axes}&&Orthogonal arms (data axes)&&\o{}&\o{}&\xo{}&\o{}&&\xs{}&\o{}&\o{}\vspace{2pt}\tabularnewline

34.&CoDa&2020&\cite{veldhuis2020coda}&&Tokens (data points)&&\xd{}&\o{}&\o{}&\o{}&&\xs{}&\o{}&\o{}\tabularnewline
&&&&&Interactive surface&&\o{}&\o{}&\o{}&\xc{}&&\xs{}&\o{}&\o{}\tabularnewline
&&&&&Sidebar buttons (filter and analytic functions)&&\o{}&\xt{}&\o{}&\o{}&&\xs{}&\o{}&\o{}\tabularnewline
&&&&&Data points&&\xd{}&\o{}&\o{}&\o{}&&\o{}&\o{}&\xl{}\tabularnewline
&&&&&Analytical functions (lines)&&\xd{}&\o{}&\o{}&\o{}&&\o{}&\o{}&\xl{}\vspace{2pt}\tabularnewline

35.&SABLIER&2022&\cite{mahieux2022sablier}&&Hourglass&&\o{}&\o{}&\xo{}&\o{}&&\xs{}&\o{}&\o{}\vspace{2pt}\tabularnewline
 
    \midrule
    
  \end{tabular}\\
    \scriptsize{\legend}
}
\caption{\captionIII}~\label{tab::illustration3}
\vspace{-9pt}
\end{table*}

\begin{table}
{\centering\small
  \begin{tabular}{@{\hspace{1pt}}r@{\hspace{3pt}}l@{\hspace{3pt}}l@{\hspace{3pt}}c}
\midrule
\textit{\#} & \textit{Application} & \textit{Domain} & \textit{Class}\tabularnewline
\midrule

1.&Slot Machine&Programming&II\tabularnewline
2.&CAAD 3D Modelling System&Architecture&II\tabularnewline
3.&Self-Builder Model (Segal Model)&Architecture&II\tabularnewline
4.&Marble Answering Machine&Telecommunication&II\tabularnewline
5.&Head Prop&Medicine (Surgery)&II\tabularnewline
6.&GraspDraw&Drawing&III\tabularnewline
7.&Tangible Geospace&Geography&II\tabularnewline
8.&BUILD-IT&Architecture&III\tabularnewline
9.&Pinwheels&Communication&I\tabularnewline
10.&mediaBlocks&Ubiquitous Computing&IV\tabularnewline
11.&musicBottles&Music Listening&II\tabularnewline
12.&Urp (Urban Planning Workbench)&Town planning&II\tabularnewline
13.&Senseboard&Scheduling&II\tabularnewline
14.&Illuminating Clay&Town planning&II\tabularnewline
15.&Audiopad&Live music synthesizing&II\tabularnewline
16.&reacTable&Live music synthesizing&II\tabularnewline
17.&IP Network Design Workbench&Information Technology&III\tabularnewline
18.&Query Shapes&3D Shapes Databases&II\tabularnewline
19.&TUISTER&Ubiquitous Computing&IV\tabularnewline
20.&I/O Brush&Painting&III\tabularnewline
21.&PICO&Town planning&II\tabularnewline
22.&Augmented Chemistry&Organic Chemistry&II\tabularnewline
23.&Nimio&Group Awareness&I\tabularnewline
24.&ArcheoTUI&Archaeology&II\tabularnewline
25.&GeoTUI&Geosciences&III\tabularnewline
26.&Slurp&Ubiquitous Computing&IV\tabularnewline
27.&Relief&Topography&II\tabularnewline
28.&Bidirectional mixiTUI&Live music rearranging&II\tabularnewline
29.&Teegi&Brain activity&II\tabularnewline
30.&SoundFORMS&Live music synthesizing&II\tabularnewline
31.&CairnFORM&Sustainability&I\tabularnewline
32.&reSpire&Health (Mindfulness)&I\tabularnewline
33.&Embodied Axes&Augmented Reality Spaces&IV\tabularnewline
34.&CoDa&Mathematics&II\tabularnewline
35.&SABLIER&Archaeology&IV\tabularnewline
 
\midrule
  \end{tabular}
}
\caption{Application domains and bodiness classes of the \numapps{} applications.}~\label{tab::by-classes}
\vspace{-9pt}
\end{table}

\subsection{Taxonomy Coverage}\label{sec-coverage}

\begin{figure}\centering
  \ifTWOCOLS
    \includegraphics[width=0.8\linewidth]{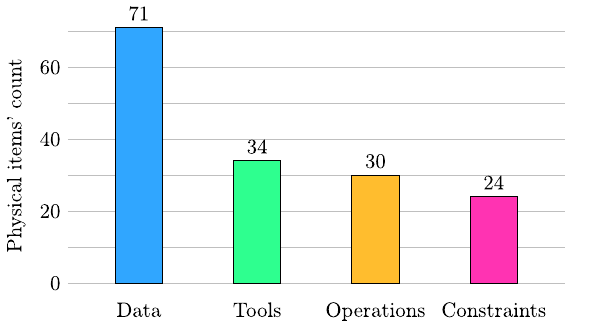} \else
    \includegraphics[width=0.4\linewidth]{tikz/histo_roles_coul.pdf} \fi
  \caption{Distribution through the four roles of the \numreps{} physical items from the collection.}
  \label{fig:roles-distribution}
  \Description{This bar chart shows the distribution of the \numreps{} physical items from the collection through the four roles of the interaction model: 71 relate to data, 34 to tools, 30 to operations, and 24 to constraints.}
\end{figure}

All the four roles are valuable in describing the \numreps{} physical items required by the \numapps{} applications. In addition, the four roles proved to be adequate in capturing all of these physical items. As illustrated in \autoref{fig:roles-distribution}, the role of data was utilized the most (\purple{71} items), followed by tools (\purple{34} items), operations (\purple{30} items), and constraints (\purple{24} items).

However, only \purple{ten} of the twelve taxonomy categories are utilized: dynamic-bodied items are absent from the operations and constraints roles of the collection; dynamic-bodied items are found only for data and tools roles (see \autoref{fig:body-per-role}). Moreover, data are mainly implemented through bodiless items, whereas tools, operations, and constraints are mainly implemented through static-bodied items.

\ifTWOCOLS
  \def\subW{0.495\linewidth}
\else
  \def\subW{0.245\linewidth}
\fi

\begin{figure}
  \centering
  \begin{subfigure}[b]{\subW{}}
    \includegraphics[width=\linewidth]{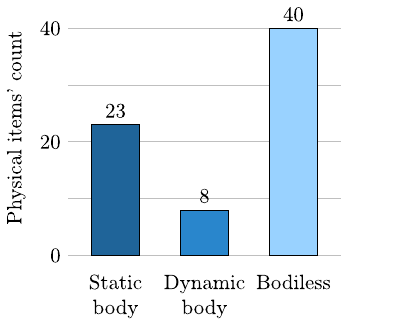} \caption{Data}
    \Description{This bar chart shows the distribution of the 71 data items from the collection through their bodiness: 23 static-bodied, 8 dynamic-bodied, and 40 bodiless representations.}
  \end{subfigure}
  \begin{subfigure}[b]{\subW{}}
    \includegraphics[width=\linewidth]{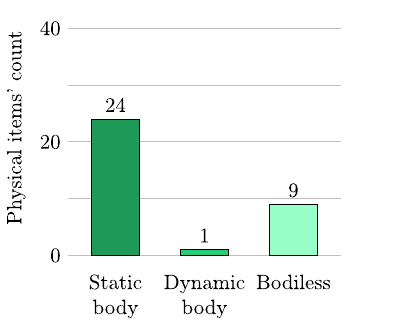} \caption{Tools}
    \Description{This bar chart shows the distribution of the 34 tool items from the collection through their bodiness: 24 static-bodied, 1 dynamic-bodied, and 9 bodiless representations.}
  \end{subfigure}
  \begin{subfigure}[b]{\subW{}}
    \includegraphics[width=\linewidth]{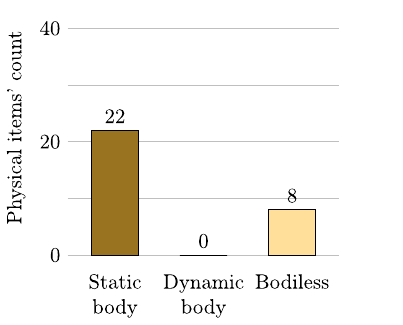} \caption{Operations}
    \Description{This bar chart shows the distribution of the 30 operation items from the collection through their bodiness: 22 static-bodied, 0 dynamic-bodied, and 8 bodiless representations.}
  \end{subfigure}
  \begin{subfigure}[b]{\subW{}}
    \includegraphics[width=\linewidth]{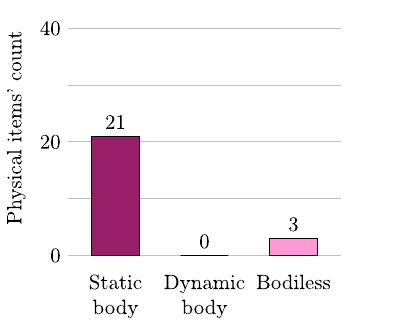} \caption{Constraints}
    \Description{This bar chart shows the distribution of the 24 constraint items from the collection through their bodiness: 21 static-bodied, 0 dynamic-bodied, and 3 bodiless representations.}
  \end{subfigure}
  \caption{Bodiness distribution across the four roles in the \numreps{} physical items from the collection of applications.}
  \label{fig:body-per-role}
\end{figure}

\subsection{Class Distribution Over Time}\label{sec-over-time}

\begin{figure}\centering
\ifTWOCOLS
    \includegraphics[trim=2.4cm 2.2cm 3.1cm 0.2cm, clip,width=\linewidth]{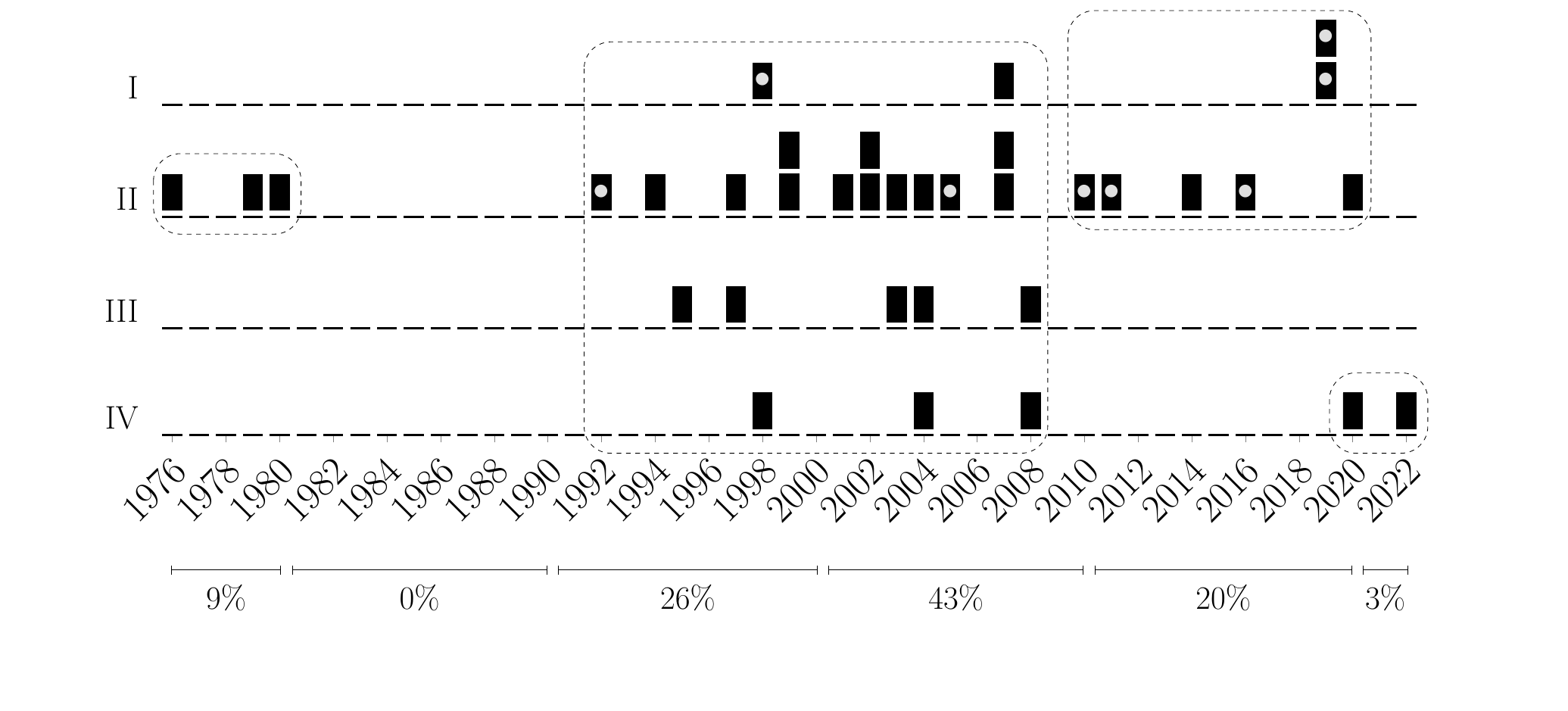}\\ \else
    \includegraphics[trim=2.4cm 2.2cm 3.1cm 0.2cm, clip,width=0.5\linewidth]{tikz/histo_classes.pdf}\\ \fi
  \scriptsize{\ita{Note.} Specimens marked with a circle utilize dynamic-bodied items.\linebreak Percentages do not total 100 due to rounding.}
  \caption{Distribution of the \numapps{} specimens, by class and time.}
  \Description{This diagram shows four superimposed timelines spanning from 1976 to 2022, each pertaining to a specific application class. Applications are clustered into four phases. The first phase, spanning from 1976 to 1980, comprises three applications from Class II. The second phase, spanning from 1992 to 2008, comprises twenty-three applications distributed across the four classes. The third phase, spanning from 2010 to 2020, comprises seven Class I and Class II applications. Finally, the fourth phase, from spanning 2020 to 2022, comprises two applications from Class IV. Only applications from Class I and Class II of the second and third phases utilize dynamic-bodied representations.}
  \label{fig:classes-distribution-over-time}
\end{figure}

The application's distribution over time through the four classes provides insights into the research phases in the field of tangible interfaces. To this end, \autoref{fig:classes-distribution-over-time} depicts and clusters the distribution of the \numapps{} applications from the collection by class and time.
A highly productive phase spans over fifteen years from the mid-1990s to the late 2000s, with \purple{66\%} of the applications ranging from 1992 to 2008 (\purple{23 out of \numapps{}}). All four classes have applications during this phase. This abundance must align with a research endeavor to prove the utility of the novel technologies introduced by the emergence of the tangible interface concept.
At present, the exploration of Class~III applications appears to remain limited to this phase, with the majority of specimens concentrated on examining tangible tabletops (\purple{four} out of \purple{five} applications involved horizontal surfaces). Subsequently, the frequency of applications decreased during a less productive phase in the 2010s due to increased emphasis on developing novel technologies for actuated and shape-changing tangible interfaces \cite{ishii2012radical,poupyrev2007actuation,rasmussen2012sci}.
Accordingly, the resulting more scarce applications, concentrated on classes~I and~II, appear to be more focused on finding applications for actuation and shape change. From 2010, applications resorting to dynamic-bodied items account for \purple{five} out of \purple{seven}, whereas they account for only \purple{three} out of \purple{twenty-three} from 1992 to 2008.
Finally, in the early 2020s, exploration of Class~IV applications resumed after \purple{a decade-long} hiatus, with a last phase comprising two specimens implementing operations in augmented reality and virtual reality scopes (\ie{} Embodied Axes \cite{cordeil2020axes}(\numaxes{}) and SABLIER \cite{mahieux2022sablier}(\numsablier{}), respectively).

\subsection{Item Distribution Over Classes}\label{sec-over-classes}

Analyzing item's bodiness give some insights into the design trends among the four classes. To this end, \autoref{tab::res-distribution-over-classes} presents the utilization of item kinds in the \numapps{} applications from the collection, categorized by bodiness class.

Tools are absent from the \purple{four} Class~I applications, which focus on dynamic-bodied data that are noneditable, either by implementing a tool or by itself. Nimio items \cite{brewer2007nimio}(\numnimio{}) can sense input on their own, but are static-bodied data. In contrast, Class~II applications appear to frequently implement tools; bodiless data items must amplify the presence of tools, as also observed in Class~III applications.
Likewise, operations are absent from Class~I applications but are frequent in Class~II and Class~III applications. The increased presence of tools and operations within Class~II and Class~III applications indicates richer interaction and more complete applications compared to Class~I.
Most tools, operations, and constraints consist of static-bodied or bodiless items; tools made of dynamic-bodied items are utilized by only \purple{one} of the \numapps{} applications (\ie{} bidirectional mixiTUI \cite{pedersen2009mixitui,pedersen2011tangiblebots}(\nummixitui{})).

\begin{table}
{\centering\small
\begin{tabular}{lllllll}
\midrule
\multicolumn{2}{l}{\textbf{\textit{Physical Item Kind}}} && \multicolumn{4}{l}{\textbf{\textit{Bodiness Class}}} \tabularnewline
\cmidrule{1-2}\cmidrule{4-7}
\textit{Role} & \textit{Bodiness} && \textit{Class I} & \textit{Class II} & \textit{Class III} & \textit{Class IV} \tabularnewline
\midrule
 Data & Static &&\raggedright $\shortmid$ &\raggedright $\shortmid$$\shortmid$$\shortmid$$\shortmid$$\shortmid$$\shortmid$$\shortmid$$\shortmid$$\shortmid$$\shortmid$$\shortmid$$\shortmid$$\shortmid$$\shortmid$$\shortmid$$\shortmid$ &\raggedright $\varnothing$ &\raggedright $\varnothing$\tabularnewline 
  & Dynamic &&\raggedright $\shortmid$$\shortmid$$\shortmid$ &\raggedright $\shortmid$$\shortmid$$\shortmid$$\shortmid$$\shortmid$ &\raggedright $\varnothing$ &\raggedright $\varnothing$\tabularnewline 
  & Bodiless &&\raggedright $\varnothing$ &\raggedright $\shortmid$$\shortmid$$\shortmid$$\shortmid$$\shortmid$$\shortmid$$\shortmid$$\shortmid$$\shortmid$$\shortmid$$\shortmid$$\shortmid$$\shortmid$$\shortmid$$\shortmid$$\shortmid$$\shortmid$$\shortmid$$\shortmid$$\shortmid$$\shortmid$ &\raggedright $\shortmid$$\shortmid$$\shortmid$$\shortmid$$\shortmid$ &\raggedright $\varnothing$\vspace{2pt}\tabularnewline 
 Tools & Static &&\raggedright $\circ$ &\raggedright $\shortmid$$\shortmid$$\shortmid$$\shortmid$$\shortmid$$\shortmid$$\shortmid$$\shortmid$$\shortmid$ &\raggedright $\shortmid$$\shortmid$$\shortmid$$\shortmid$$\shortmid$ &\raggedright $\varnothing$\tabularnewline 
  & Dynamic &&\raggedright $\circ$ &\raggedright $\shortmid$ &\raggedright $\circ$ &\raggedright $\varnothing$\tabularnewline 
  & Bodiless &&\raggedright $\circ$ &\raggedright $\shortmid$$\shortmid$$\shortmid$ &\raggedright $\shortmid$$\shortmid$ &\raggedright $\varnothing$\vspace{2pt}\tabularnewline 
 Operations & Static &&\raggedright $\circ$ &\raggedright $\shortmid$$\shortmid$$\shortmid$$\shortmid$$\shortmid$$\shortmid$$\shortmid$$\shortmid$$\shortmid$$\shortmid$ &\raggedright $\shortmid$$\shortmid$$\shortmid$ &\raggedright $\shortmid$$\shortmid$$\shortmid$$\shortmid$$\shortmid$\tabularnewline 
  & Dynamic &&\raggedright $\circ$ &\raggedright $\circ$ &\raggedright $\circ$ &\raggedright $\circ$\tabularnewline 
  & Bodiless &&\raggedright $\circ$ &\raggedright $\shortmid$$\shortmid$$\shortmid$ &\raggedright $\shortmid$$\shortmid$ &\raggedright $\circ$\vspace{2pt}\tabularnewline 
 Constraints & Static &&\raggedright $\circ$ &\raggedright $\shortmid$$\shortmid$$\shortmid$$\shortmid$$\shortmid$$\shortmid$$\shortmid$$\shortmid$$\shortmid$$\shortmid$$\shortmid$$\shortmid$$\shortmid$ &\raggedright $\shortmid$$\shortmid$$\shortmid$ &\raggedright $\shortmid$\tabularnewline 
  & Dynamic &&\raggedright $\circ$ &\raggedright $\circ$ &\raggedright $\circ$ &\raggedright $\circ$\tabularnewline 
  & Bodiless &&\raggedright $\circ$ &\raggedright $\shortmid$$\shortmid$ &\raggedright $\shortmid$ &\raggedright $\circ$\vspace{2pt}\tabularnewline 
\midrule
\end{tabular}

   \scriptsize{
    \ita{Note.} ``$\shortmid$'' = one application using at least one of item of this kind. ``$\circ$'' = no application uses an item of this kind. ``$\varnothing$'' = this item kind is unavailable in this class.\\
  }
}
\caption{Use of item kinds by class in the \numapps{} applications.}~\label{tab::res-distribution-over-classes}

\end{table}

Thereby, the taxonomy and the classification afford the accurate framing and detailed analysis of \numapps{} applications from the literature and their \numreps{} items, and reveal some design opportunities that remain to explore. The next section offers perspectives on the field.

\section{Discussion}\label{sec-discussion}

This section provides insights into the evolution of the field and the future exploration of tangible interface design. Additionally, it addresses some limitations of the present analysis.

\subsection{Evolution of the Research Field}\label{sec-discussion-evolution}

The main effort in finding new applications for the tangible interface concept slowed down fifteen years ago. Since then, recent efforts have focused on taking advantage of dynamic-bodied items. Nevertheless, the scarcity of applications since 2010, compared to the two preceding decades, suggests that the community could now be more active in this pursuit. Potential challenges that hinder progress involve expertise, finances, and time:

\begin{enumerate}[label=(\alph*),ref=\alph*]
  
\item Developing tangible interface applications with dynamic-bodied items requires HCI students and researchers to integrate additional skills beyond computer programming, physical prototyping, and user-centered design. They must become more knowledgeable about sensing and actuating possibilities \cite{boer2021smorgasbords} and deepen their collaboration with engineers from mechanics and mechatronics.

\item Physical computing is more expensive than graphical computing \cite{holmquist2023cheap}, especially when actuation capabilities are required. For instance, whereas a small 12$\times$12 array of actuated pins already necessitates 144 motors, doubling lines' lenght to get a 24$\times$24 array, like the one in SoundFORMS \cite{colter2016soundforms}(\numsoundform{}), balloons to 576 motors, thus increasing cost by 4.

\item The same, the development time of such dynamic interfaces increases as the intricacy grows very quickly, for example, according to the number of spare parts to produce, store, and assemble, and the number of cable of the wiring to cut, prepare, and mount. As well, writing the driver for new hardware adds to the applications' development time. All these steps render design and prototyping iterations even more challenging.

\end{enumerate}

This way, the exploration of technical solutions to reduce the number of components and development time is still pending. Such solutions will not only contribute to decreasing the environmental impact but also improve economic viability \cite{holmquist1999token,holmquist2023cheap}. Only then, researchers could intensify the exploration of Class~I and Class~II applications and resume the exploration of Class~III applications.
To sustain this effort, three levers are necessary:
\begin{enumerate}
\item Breaking the dependence on manufactured bodied items (\eg{} liberated pixels \cite{holmquist2023cheap}, constructive inputs \cite{liang2014gaussbricks,zheng2019mechamagnets}, and molecular user interfaces \cite{pruszko2021molecular,pruszko2023modular});
\item Reducing the technological cost and complexity of actuation and shape-change mechanisms \cite{alexander2018challenges,jansen2015opportunities}, for example, by investing the low tech aspiration \cite{bihouix2020age,mirmalek2017lowtech};
\item Encouraging publications that illustrate design concepts of such applications---even if only through sketches, simulations, and interaction vocabularies \cite{diefenbach2013vocabulary,hornecker2023vocabulary,rasmussen2016sketching}---can provide valuable guidance for further technological investigations.
\end{enumerate}

Additionally, the increased financial and environmental costs associated with the development of bespoke hardware for tangible interfaces \cite{holmquist2019future,holmquist2023cheap} compared to graphical interfaces suggest to not only focus on the embodiment of new data \cite{holmquist2023cheap}. A promising path lies in the adjunction of meaningful bodied tools and operations to pixel-based interfaces other than tangible interfaces \cite{holmquist2019future,holmquist2023cheap}. The recent resurgence of Class~IV exploration embraces this emerging trend by adding bodied operations to two augmented reality and virtual reality applications (\ie{} Embodied Axes \cite{cordeil2020axes}(\numaxes{}) and SABLIER \cite{mahieux2022sablier}(\numsablier{}), respectively). A third recent example in 2019 is an old-fashioned hotel bell for front desks that allows guests to call a remote receptionist who is viewed on a monitor \cite{holmquist2019future}. Such examples of Class~IV applications still require further development, and should benefit from collaboration with researchers who focus on other interface scopes such as graphical interfaces, virtual reality, and augmented reality.

\subsection{Perspectives for Design}\label{sec-discussion-design}

Surprisingly, Class~I applications appear to be infrequent, despite being the most aligned with the original ``tangible bits'' vision \cite{ishii1997tangible}. Additionally, these applications are predominantly composed of noneditable dynamic-bodied data. For instance, editable dynamic-bodied data, including shape-changing items---which require more extensive integration of user input capabilities \cite{rasmussen2012sci}---are still untapped in Class~I applications. To design richer interaction in such applications, future work must investigate self-editable dynamic-bodied data or combine dynamic-bodied data with bodied tools. In addition, constraints are primarily utilized in Class~II and Class~III applications. Investigating the use of constraints can also aid in the design of more sophisticated interactions in Class~I and Class~IV applications.

The implementation of dynamic-bodied tools, operations, and constraints should be further explored to potentially improve application in all of the four bodiness classes. Technology is ready, such as shape-changing buttons \cite{tiab2016understanding}, expandable dials \cite{kim2019expandial}, and expandable pucks \cite{kim2018knobslider,oosterhout2020reshaping}, that need to benefit to Class~II and Class~III applications. For example, integrating active pucks \cite{pedersen2011tangiblebots} into mixiTUI \cite{pedersen2009mixitui} enhances interaction by providing haptic feedback for use ranges (\ie{} to keep between minimum and maximum values), predefining computer-generated puck rotations (to play pre-composed effects), and grouping pucks for rotation imitation (to synchronize two audio tracks). Moreover, dynamic-bodied constraints still need to be explored. As an illustration, AeroRigUI \cite{lilith2023aerorigui} proposes a movable surface that can serve, when appropriate, as a separation or a projection screen. Another illustration is simulating the Marble Answering Machine with inFORM by dynamically shaping the constraints of the tangible user interface thanks to the array of pins \cite{follmer2013inform}. This way, dynamic constraints enable switching between applications and tasks, which is also a promising direction for the commercial viability of tangible interfaces.

\subsection{Limitations of the Analysis}\label{sec-limitations-future-work}

A limitation of this work is the use of a taxonomy to describe the specimens without consultation with their designers. 
Describing the \numapps{} applications was relatively straightforward, but at times, advanced questioning was required to comprehend their designers' thinking. For instance, the Pinwheels specimen \cite{wisneski1998pinwheels}(\numpinwheels{}), which represents information flow ambiently, can be interpreted from two viewpoints. At first glance, Pinwheels appear to consist of bodiless data items, where airflow represents the data flow. However, this idea has evolved from using airflow in the ambientROOM \cite{ishii1998ambientroom}, simply requiring air to enter through a hole, to using ``spinning pinwheels'' \cite{dahley1998ambient,ishii2001pinwheels,wisneski1998pinwheels} as a visual representation. As a result, Pinwheels are now referred to as bodied data items. 

Another example regards the inclusion of CairnFORM \cite{daniel2019cairnform}(\numcairnform{}) in Class~I, but the exclusion of Relief \cite{leithinger2010relief,leithinger2011relief}(\numrelief{}). CairnFORM utilizes bodied rings to convey data through a combination of shape and color changes. Because color is determined by the physical properties of the rings, CairnFORM's energy forecast was considered as a Class~I application. Relief also combines shape and color changes of a bodied item to convey a topographic map. However, because the map display exceeds the boundaries of the bodied shape, topography encoding with Relief was considered a Class~II application. Without this factor, Relief would have fallen into Class~I. Therefore, knowing whether the overflow is a deliberate design choice, as seen in overview+detail or focus+context interfaces, holds implications for classification considerations. Nevertheless, despite being essential for generalization and standardization, these classification considerations remain secondary to the prevailing concerns of user experience and performance, which must primarily guide user interface design.

Therefore, advanced questioning can become challenging, as design thoughts are not always detailed in the publications. The community should benefit from future work that dissect prior specimens in a publicly accessible database (\eg{} \cite{fleck2018cladistics}), and from future publications that clearly describe items' bodiness and the roles in which they are involved.

\section{Conclusion}\label{sec-conclusion}

This article confirms four distinct phases in the emergence of tangible interface applications: spanning from the premises laid in the late 1970s and early 1980s to the productive exemplification of the concept in the 1990s and 2000s, the tight integration of actuation and shape-changing technologies in the 2010s, and the emerging integration in pixel-based interfaces in the early 2020s. A taxonomy outlining the physical items present in applicative tangible interface enhances understanding of the four phases and uncovers overlooked possibilities. The taxonomy consists of two axes: the roles of applications' interactive components and the bodiness of the physical items involved in these components (considering that bodiless items---such as graphics and sound---are physical). The four roles of data, tools, operations, and constraints, which are articulated within a new interaction model for applicative tangible interfaces, accurately discriminated the \numreps{} physical items from a collection of \numapps{} applications from the literature. Building upon the taxonomy, four bodiness classes are identified according to how the data role is implemented.

Three potential paths for future tangible interface applications are promising. (a) The first path is to broaden the scope of Class~I applications to editable dynamic-bodied data items through the implementation of self-editable data or the addition of bodied tools. (b) The second path is to gain leverage of dynamic-bodied tools, operations, and constraints that are currently underrepresented in applicative tangible interfaces despite mature technology. This effort must directly benefit to applications from Classes~I, II, and~III. Three drivers support this effort: (1) breaking the dependence on manufactured bodied items, (2) reducing the technological cost and complexity of actuation and shape change mechanisms, and (3) encouraging publications that illustrate the design concepts of such applications, even if only through sketches, simulations, and interaction vocabularies. (c) Finally, the third path is to bring tangibility to pixel-based interfaces by implementing meaningful bodied operations from Class~IV applications in scopes beyond tangible interfaces, including graphical interfaces, augmented reality, and virtual reality. Advancing along these \purple{three} paths ought to contribute to the fulfillment of the Tangible Bits vision and get tangible interfaces a step closer to their commercial availability and viability.

\bibliographystyle{ACM-Reference-Format}
\bibliography{paper.bib}

\end{document}